%% file: main.tex
\documentclass[conference,letterpaper]{IEEEtran}
\IEEEoverridecommandlockouts

\usepackage[utf8]{inputenc} 
\usepackage[T1]{fontenc}
\usepackage{url}
\usepackage{ifthen}
\usepackage{cite}
\usepackage[cmex10]{amsmath} % Use the [cmex10] option to ensure complicance
\PassOptionsToPackage{hyphens}{url}
\usepackage[hidelinks]{hyperref}
\usepackage{vmr-symbols-vecbold}
\usepackage{standard-macros}
\usepackage{algorithm}
\usepackage{algpseudocode}
\usepackage{cuted}

\usepackage{tikz,pgfplots}
\usepackage{pgfplotstable}
\usetikzlibrary{positioning}
\usetikzlibrary{decorations.pathreplacing}
\usetikzlibrary {arrows.meta,bending,positioning}
\pgfplotsset{compat=newest}
\usepgfplotslibrary{patchplots}
\usepgfplotslibrary{statistics}
\usetikzlibrary{backgrounds,plotmarks,fit,positioning,arrows,shapes,shapes.multipart,calc,arrows.meta}
\usetikzlibrary{fit,positioning,arrows,shapes,shapes.multipart,calc,arrows.meta,shapes.geometric,shapes.misc}
\usetikzlibrary{decorations.pathreplacing,angles,quotes,calligraphy}
\usetikzlibrary{automata}
\usetikzlibrary{arrows.meta, positioning, backgrounds, fit}
\usepackage{enumitem}
\usepackage{adjustbox}
\usepackage{mathtools}
\usepackage{siunitx}

\usepackage[hidelinks]{hyperref}

\DeclareMathOperator{\Bin}{Bin}

\DeclareMathOperator{\Unif}{Unif}
\DeclareMathOperator{\enc}{enc}
\DeclareMathOperator{\dec}{dec}
\newcommand{\adjustedbar}[1]{\overline{\hspace{-0.1em}#1\hspace{-0.1em}}}
\newcommand{\dTV}{\ensuremath{\overline{\opT\opV}}}

\begin{document}
\title{
Type-Based Unsourced Multiple Access over Fading Channels with Cell-Free Massive MIMO
}

% %%% Single author, or several authors with same affiliation:
% \author{%
%  \IEEEauthorblockN{Author 1 and Author 2}
% \IEEEauthorblockA{Department of Statistics and Data Science\\
%                    University 1\\
 %                   City 1\\
  %                  Email: author1@university1.edu}% }

%%% Several authors with up to three affiliations:
\author{
\IEEEauthorblockN{Kaan Okumus$^{*}$, Khac-Hoang Ngo$^{\dagger}$, Giuseppe Durisi$^{*}$, and Erik G. Str\"om$^{*}$
}
\IEEEauthorblockA{
$^{*}$Department of Electrical Engineering, Chalmers University of Technology, 41296 Gothenburg, Sweden\\ Email: \{okumus, durisi, erik.strom\}@chalmers.se\\
$^{\dagger}$Department of Electrical Engineering, Link\"oping University, 58183 Link\"oping, Sweden\\ Email: khac-hoang.ngo@liu.se 
\thanks{This work was supported in part by the Swedish Research Council under grants 2021-04970 and 2022-04471, and by the Swedish Foundation for Strategic Research. The work of K.-H. Ngo was supported in part by the Excellence Center at Linköping – Lund in Information Technology (ELLIIT).
}
}
%
%  \IEEEauthorblockN{Author 1}
%  \IEEEauthorblockA{Department of Electrical Engineering \\
%                    University 1\\
%                    City 1\\
%                    Email: author1@university1.edu}
%  \and
%  \IEEEauthorblockN{Author 2 and Author 3}
%  \IEEEauthorblockA{Research Center XY\\ 
%                    City 2\\
%                    Email: \{author2, author3\}@research-center.com}
}

\maketitle

%%%%%%
%% Abstract: 
%% If your paper is eligible for the student paper award, please add
%% the comment "THIS PAPER IS ELIGIBLE FOR THE STUDENT PAPER
%% AWARD." as a first line in the abstract. 
%% For the final version of the accepted paper, please do not forget
%% to remove this comment!
%%

\begin{abstract} 
Type-based unsourced multiple access (TUMA) is a recently proposed framework for type-based estimation in massive uncoordinated access networks. We extend the existing design of TUMA, developed for an additive white Gaussian channel, to a more realistic environment with fading and multiple antennas. Specifically, we consider a cell-free massive multiple-input multiple-output system and exploit spatial diversity to estimate the set of transmitted messages and the number of users transmitting each message. Our solution relies on a location-based codeword partition and on the use at the receiver of a multisource approximate message passing algorithm in both centralized and distributed implementations. The proposed TUMA framework results in a robust and scalable architecture for massive machine-type communications. 
\end{abstract}

\section{Introduction}

Massive machine-type communication is pivotal for the internet of things, enabling connectivity for a massive number of devices (also called users). These devices, ranging from sensors to smart appliances, demand scalable and energy-efficient communication systems to support high-density deployments. Unsourced multiple access (UMA), introduced by Polyanskiy~\cite{polyanskiy2017}, provides a theoretical framework for the analysis of massive uncoordinated access systems. In UMA, all devices use the same codebook, and the receiver decodes the set of transmitted messages without identifying their sources. In both the original analysis~\cite{polyanskiy2017} and many follow-up extensions~\cite{amalladine_ccs_amp, fengler_nonbayesian, fengler_sparc, ngo_2023_random_user_activity}, the event where multiple devices transmit the same message simultaneously, referred to as message collisions, is treated as error. Indeed, under the assumption that each device chooses its message uniformly at random from a large set, the probability that two devices pick the same message is negligible. However, there are many practically relevant scenarios, such as industrial monitoring, multi-target tracking~\cite{Hoffman2004}, point-cloud transmission~\cite{Bian24wpc}, and federated learning~\cite{qiao_gunduz_fl}, in which the messages are related to underlying physical or digital processes and, hence, may be correlated. In such scenarios, it is often necessary for the receiver not only to decode the message set, but also to estimate multiplicities, i.e., the number of users transmitting the same message.

The idea of estimating the type, i.e., the empirical distribution of messages across the users, dates back to the work of Mergen and Tong~\cite{Mergen2006}. Type-based UMA (TUMA), introduced by Ngo \textit{et al.}~\cite{ngo2024_tuma}, extends the approach in~\cite{Mergen2006} to the UMA framework by letting the receiver decode the set of transmitted messages along with their multiplicities. In~\cite{ngo2024_tuma}, TUMA was designed and validated for an additive white Gaussian noise (AWGN) channel under perfect power control; in this simplified scenario, multiplicities can be estimated directly from the power at which each codeword is received.

The purpose of this paper is to generalize the analysis in \cite{ngo2024_tuma} to the case of fading channels. Specifically, we shall consider TUMA over a cell-free (CF) massive multiple-input multiple-output (MIMO) system~\cite{cf_mimo_book}. We choose this architecture to leverage the benefits of distributed connectivity~\cite{cf_mimo_book} in type estimation. Gkiouzepi \textit{et al.}~\cite{gkiouzepi2024jointmessagedetectionchannel} recently demonstrated the benefits of CF massive MIMO for UMA using the multisource approximate message passing (AMP) algorithm proposed in~\cite{cakmak_2025_journal}. Specifically, they showed that multisource AMP combined with location-based codeword partition in an UMA setting allows not only for message recovery, but also for the accurate estimation of the position of each device. However, their framework does not account for message collisions. AMP-based digital aggregation (AMP-DA),  introduced in the federated learning framework by Qiao \textit{et al.}~\cite{qiao_gunduz_fl}, addresses collisions and mitigates fading via channel pre-equalization at the transmitter. However, this approach requires perfect channel state information (CSI) at the devices, which is impractical because it is onerous to acquire. 

In this paper, we show that the same two main tools used in \cite{gkiouzepi2024jointmessagedetectionchannel}, namely, location-based codeword partition and multisource AMP, allow for type estimation in a TUMA system operating over a CF massive MIMO architecture, without the need for CSI at the devices or the receiver. We also illustrate that satisfactory performance can be achieved when a centralized decoder is replaced by a scalable distributed decoder inspired by the distributed AMP (dAMP) algorithm~\cite{bai_larsson_damp}. Numerical results demonstrate that the proposed decoders outperform AMP-DA in estimating the type when CSI at the devices in AMP-DA is imperfect.

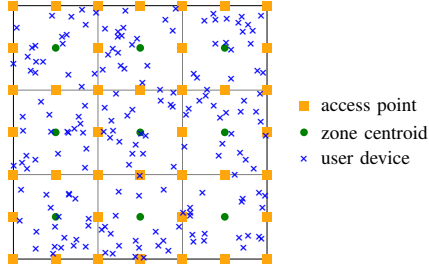
\begin{figure}[t!]
    \centering
    \resizebox{0.65\columnwidth}{!}{\input{figs/cf_topology}}
    %\vspace{-0.22cm}
    \caption{An example topology of the proposed TUMA framework within a CF massive MIMO network over $\uU=9$ zones.}
    \label{fig:topology}
    \vspace{-0.4cm}
\end{figure}

\subsubsection*{Notation}

System parameters are denoted by uppercase nonitalic letters (e.g., $\uA$), 
sets by calligraphic letters (e.g., $\setS$),
vectors by bold italic lowercase letters (e.g., $\vecx$), 
and matrices by bold nonitalic uppercase letters (e.g., $\mathbf{X}$). 
We write the element on the $a$th row and $b$th column of $\mathbf{X}$ as $[\mathbf{X}]_{a,b}$, and the $b$th element of $\vecx$ as~$[\vecx]_b$. We denote the $n \times n$ identity matrix by $\mathbf{I}_n$, and transposition and Hermitian transposition by $\tp{}$ and~$\herm{}$, respectively. We denote the complex proper Gaussian vector distribution with mean $\veczero$ and covariance $\mathbf{A}$ by $\mathcal{C}\mathcal{N}(\veczero, \mathbf{A})$, and its probability density function by $\mathcal{C}\mathcal{N}(\cdot; \veczero, \mathbf{A})$. Uniform distribution over~$(a,b)$ is denoted by $\Unif(a,b)$, $\ell_p$-norm by $\vecnorm{\cdot}_p$; $\Re(\cdot)$ returns the real part and $[n]\coloneqq\{1,\dots,n\}$. We denote the Kronecker delta function by $\delta(\cdot)$, the Kronecker product by $\otimes$, and elementwise multiplication by $\odot$; $\diag(x_1, \cdots, x_n)$ is a diagonal matrix with $x_1, \dots, x_n$ as its diagonal entries. The notation~$\sim_{\text{i.i.d.}}$ indicates independent and identically distributed (i.i.d.) elements from the specified distribution. Finally, we denote the probability simplex over the set $[\uM]$ by $\mathcal{P}([\uM])$.

\section{System Model}\label{Sec:SystemModel}

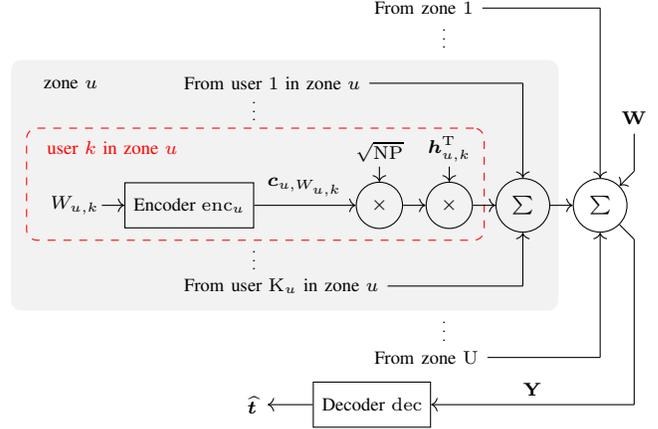
\begin{figure}[t!]
    \centering
    \vspace{-0.1cm}
    \input{figs/block_diagram}
    %\vspace{-0.4cm} % Adjust this value as needed
    \caption{Block diagram of the proposed TUMA framework with fading channel in a CF system.}
    \label{fig:block-diagram-squeezed}
    \vspace{-0.5cm}
\end{figure}

We consider a CF system where $\uB$ access points (APs) are connected to a central processing unit (CPU) via fronthaul links. The APs collaboratively serve single-antenna users, randomly located in a coverage area $\setD$. The area is partitioned into $\uU$ nonoverlapping zones $\{\setD_u\}_{u=1}^\uU$, such that $\setD = \bigcup_{u=1}^{\uU} \setD_u$ and $\setD_u \cap \setD_{u'} = \emptyset$, $\forall u \neq u'$. Each AP~$b$ is located at position $\nu_b \in \setD$ and equipped with $\uA$ antennas, yielding $\uF = \uA \times \uB$ antennas in total. We~illustrate~an~example of system topology in Fig.~\ref{fig:topology}, where the area is divided~into~a $3\times 3$ grid of square zones, with APs~evenly~placed~along the zone boundaries. While our model and design are applicable~to general topology, we use this specific topology in the simulations in Section~\ref{Sec:Sim} for its ability to ensure uniform coverage of the area. The overall operation of the proposed TUMA framework, encompassing message encoding, transmission through the fading channel, and decoding at the receiver, is summarized in Fig.~\ref{fig:block-diagram-squeezed}. Next, we detail each block of this diagram. 

\subsection{Messages and Encoder}

Each user $k$ in zone $u$ selects a message $W_{u,k}$ from the message set $[\uM]$, where $\uM$ is the total number of possible messages. These messages might be obtained from a quantization of the user's data, which may be, for example, local updates in federated learning or targets' position in multi-target tracking.\footnote{Different from \cite{ngo2024_tuma}, where both quantization and communication are considered in the TUMA model, we focus here for simplicity only on communication, i.e., on the encoder and decoder design.} The system employs an UMA codebook $\mathbf{C} \in \mathbb{C}^{\uN \times \adjustedbar{\uM}}$, where $\uN$ is the blocklength and $\adjustedbar{\uM} = \uU \cdot \uM$ is the total number of codewords. The codebook is evenly partitioned into zone-specific subcodebooks: $\mathbf{C} = [\mathbf{C}_1, \cdots, \mathbf{C}_{\uU}]$, where $\mathbf{C}_u = [\vecc_{u,1}, \cdots, \vecc_{u,\uM}] \in \mathbb{C}^{\uN \times \uM}$, where $\Exop [\lVert \vecc_{u,m} \rVert^2_2] \leq 1$, $\forall u\in [\uU]$, $m\in[\uM]$. The set of column vectors of $\mathbf{C}_u$ forms the set of codewords for zone $u$, denoted as $\setC_u = \{\vecc_{u,1},\dots, \vecc_{u,{\uM}}\}$. The encoder $\enc_{u} \sothat [\uM] \rightarrow \setC_u$ maps each message $W_{u,k}$ to the codeword $\enc_{u}(W_{u,k}) = \vecc_{u,W_{u,k}}$. Note that all users within the same zone use the same encoder.

\subsection{Multiplicity and Type} \label{SubSec:MultType}

Let $\uK_{u}$ denote the number of active users in zone $u$ and $\uK_{\mathrm{a}}=\sum_{u=1}^\uU \uK_{u}$ denote the total number of active users. Furthermore, let $\uM_{\mathrm{a},u}$ denote the number of distinct transmitted messages in zone $u$. We denote the number of users transmitting the codeword $m$ in zone $u$ by $k_{u,m} \in \{0,1,\dots,\uK_u\}$. The $\{k_{u,m}\}$ form the multiplicity vector $\veck_u = \tp{[k_{u,1},\cdots,k_{u,{\uM}}]}$ with $ \lVert \veck_u \rVert_1 = \uK_{u}$ and $ \lVert \veck_u \rVert_0 = \uM_{\mathrm{a},u}$. The global multiplicity vector is defined as $\veck = \tp{[k_{1},\cdots,k_{{\uM}}]}$ with $k_m = \sum_{u=1}^\uU k_{u,m}$. The type is then obtained as the vector of normalized multiplicities, i.e., $\vect = \tp{[t_{1},\cdots,t_{{\uM}}]}$ with $t_{m} = k_{m}/\uK_{\mathrm{a}}$.

\subsection{Channel Model}

We index by $(u,k)$ the $k$th user in zone~$u$. The channel between user~$(u,k)$ and AP~$b$ is modeled as a quasi-static Rayleigh fading channel. Specifically, the channel coefficients are independent across antennas, APs, and users. The channel vector $\vech_{u,k} \in \mathbb{C}^{\uF}$ between user $(u,k)$ and all receive antennas is distributed as $\mathcal{C}\mathcal{N}(\veczero, \matSigma(\rho_{u,k}))$, where $\rho_{u,k}$ denotes the position of user~$(u,k)$ and $\matSigma(\rho_{u,k}) = \diag(\gamma_1(\rho_{u,k}), \cdots, \gamma_{\uB}(\rho_{u,k})) \otimes \mathbf{I}_{\uA}$ with $\gamma_b(\rho_{u,k})$ being the large-scale fading coefficient (LSFC). When multiple users transmit the same codeword, their contributions are superimposed at the receiver. For a codeword $\vecc_{u,m}$, the effective channel vector is
\begin{equation}
\vecx_{u,m} = 
    \begin{cases}
    \sum_{k=1 \sothat W_{u,k}=m}^{k_{u,m}} \vech_{u,k} & \text{if } k_{u,m}>0, \\
    \veczero & \text{if } k_{u,m}=0.
    \end{cases} \label{expression_x} 
\end{equation}
The effective channel vectors for all codewords in zone $u$ form the effective channel matrix $\mathbf{X}_u = \tp{[\vecx_{u,1}, \cdots, \vecx_{u,{\uM}}]} \in \mathbb{C}^{{\uM} \times \uF}$. The aggregated received signal across all APs is 
\begin{equation} \label{eqn-mainreceivedsignal}
\mathbf{Y} = \sqrt{\uN\uP}  \sum_{u=1}^{\uU} \mathbf{C}_u \mathbf{X}_u + \mathbf{W},
\end{equation}
where $\mathbf{W} \sim_{\text{i.i.d.}} \mathcal{C}\mathcal{N}(0,\sigma_w^2)$ is the AWGN signal. The per-symbol average transmit power is $\uP$. Therefore, the transmit signal to noise ratio (SNR) is $\text{SNR}_{\text{tx}} = \uP / \sigma_w^2$.

\subsection{Decoder} \label{Sec:SystemModel-Decoding}

The decoder estimates the message type by using a decoding function defined as $\dec \sothat \mathbb{C}^{\uN \times \uF} \rightarrow \setP([{\uM}])$. Given $\mathbf{Y}$ and $\mathbf{C}$, the decoder first estimates the multiplicities per zone, $\widehat{\veck}_u = \tp{[\widehat{k}_{u,1}, \cdots, \widehat{k}_{u,{\uM}}]}$, and then computes the global multiplicity vector, $\widehat{\veck} = \sum_{u=1}^\uU \widehat{\veck}_u$. Finally, the type is estimated as $\widehat{\vect} = \widehat{\veck} / \lVert \widehat{\veck} \rVert_1$. The performance of type estimation is evaluated using the average total variation (TV) distance between the type of the transmitted messages and its estimate, defined as
\begin{equation}
\dTV = \frac{1}{2} \Exop \left[  \sum_{m=1}^{{\uM}} | t_{m} - \widehat{t}_{m}| \right], \label{eq:tv_dist_expression}
\end{equation}
where the expectation is over the randomness of messages, types, user positions, small-scale fading, and additive noise. 

\section{Proposed Decoder} \label{Sec:Decoder}

For the introduced TUMA framework, we propose a decoder that employs the multisource AMP algorithm~\cite{gkiouzepi2024jointmessagedetectionchannel}. To handle message collisions, we consider a modified Bayesian prior and a tailored denoiser that accounts for multiplicities. Throughout, we assume that the receiver has access to the received signal, and has perfect knowledge of the codebook~$\mathbf{C}$, the LSFC model $\gamma_b(\cdot)$, the number of active users~$\uK_{u}$, and the number of unique messages $\uM_{\mathrm{a},u}$ in each zone.\footnote{The model can be extended to handle scenarios where $\uK_u$ and $\uM_{\mathrm{a},u}$ are random and unknown at the receiver. In this case, we initialize these parameters and refine their values, and also the prior, along the AMP iterations.} We first introduce a centralized decoder, then discuss approximations for efficient implementation, and finally present a scalable, distributed version of the proposed decoder.

\subsection{Centralized Decoder} \label{SubSec-CentDec}
The centralized decoder employs the multisource AMP algorithm to iteratively process the received signal and extract necessary information for multiplicity estimation.

\subsubsection{Multisource AMP} \label{subsec:multi_amp}
The algorithm performs $\uT$ iterations, where $\mathbf{X}_u^{(t)}$, the estimate of the effective channel matrix for zone $u$, and $\mathbf{Z}^{(t)}$, the residual noise, are initialized as $\mathbf{X}_u^{(0)} = \veczero$ and $\mathbf{Z}^{(0)}=\mathbf{Y}$. The updates are as follows:
\begin{subequations} \label{amp_decoder}
    \begin{align}
        &\mathbf{R}_u^{(t)} = \mathbf{C}_u^{\uH} \mathbf{Z}^{(t-1)} + \sqrt{\uN \uP}
        \mathbf{X}_u^{(t-1)}, \\
        &\mathbf{X}_u^{(t)} = \eta_{u,t}(\mathbf{R}_u^{(t)}), \label{amp_decoder_row2} \\
        &\mathbf{\Gamma}_u^{(t)} = \mathbf{C}_u \mathbf{X}_u^{(t)} - \frac{{\uM}}{\uN} \mathbf{Z}^{(t-1)} \mathbf{Q}_u^{(t)}, \label{amp_decoder_row3} \\
        &\mathbf{Z}^{(t)} = \mathbf{Y} - \sqrt{\uN \uP} 
        \sum_{u=1}^{\uU} \mathbf{\Gamma}_u^{(t)}.
    \end{align}
\end{subequations}
Here, $\mathbf{Q}_u^{(t)}$ is the Onsager term which will be characterized in Section~\ref{Subsec:onsager}. The denoiser $\eta_{u,t}(\cdot)$ operates row-wise on $\mathbf{R}_u^{(t)}$, leveraging the effective decoupled channel model
\begin{equation}
    \vecr_{u,m}^{(t)} = \sqrt{\uN\uP} 
    \vecx_{u,m} + \vecvarphi^{(t)},
    \label{decouple_model}
\end{equation}
where $\vecvarphi^{(t)} \sim \mathcal{C}\mathcal{N}(\veczero, \mathbf{T}^{(t)})$ is the effective noise and  $\mathbf{T}^{(t)}$ evolves according to state evolution, a tool for tracking the AMP algorithm's dynamics \cite{Bayati_2011}. In multisource AMP, state evolution ensures a block diagonal structure \cite{cakmak_2025_journal} for $\mathbf{T}^{(t)} = \diag(\tau_1^{(t)},\cdots, \tau_{\uB}^{(t)}) \otimes \mathbf{I}_{\uA}$, where $\tau_b^{(t)}$ is given by
\begin{equation}
    \tau_b^{(t)} = \frac{1}{\uN \uA} \sum_{a=1}^{\uA} \Re \left\{[(\mathbf{Z}^{(t-1)})^{\uH} \mathbf{Z}^{(t-1)}]_{(b-1)\uA + a, \, (b-1)\uA + a} \right\}. \label{residual_expression}
\end{equation}

\subsubsection{Prior Selection} \label{subsec:prior} An appropriate prior is essential for accurate decoding. We assume that for zone $u$, the active message set of size $\uM_{\mathrm{a},u}$ is uniformly selected from the ${\uM}$ messages. The multiplicities of the active messages are then drawn from a multinomial distribution with identical probabilities $1/\uM_{\mathrm{a},u}$, under the condition that the multiplicity is not zero. We approximate the marginal of the multinomial distribution by a binomial distribution with parameters $(\uK_u, 1/\uM_{\mathrm{a},u})$, truncated from $1$ to $\uK_u$. The resulting approximate prior is
\begin{multline}
    p(k_{u,m}=k) = p_0 \delta{(k)} \\\qquad + \left(1 - p_0\right)  \sum_{l=1}^{\uK_{u}}\frac{\Bin(l; \uK_u,1/\uM_{\mathrm{a},u}) \delta{(k-l})}{\sum_{i=1}^{\uK_{u}}\Bin(i; \uK_u,1/\uM_{\mathrm{a},u})}, \label{prior_prior_expression}
\end{multline}
where $p_0 = 1 - \uM_{\mathrm{a},u}/{\uM}$ is the probability that a message is not activated, and $\Bin(\cdot;n,p)$ denotes the binomial probability mass function with parameters $(n,p)$. Furthermore, the user positions are assumed to be independently and uniformly distributed over the coverage region $\setD_u$ for zone $u$. Given $k_{u,m}=k$, the positions of users transmitting the $m$th codeword in zone $u$ are denoted by $\vecrho_{u,m,1:k} = [\rho_{u,m,1},\dots,\rho_{u,m,k}]$. These positions follow the distribution $p(\vecrho_{u,m,1:k} \mid k_{u,m}=k) = 1/|\setD_u|^k$, where $|\setD_u|$ denotes the area of the region $\setD_u$.

\subsubsection{Denoiser} \label{subsec:denoiser} The Bayesian posterior mean estimator (PME) of $\vecx_{u,m}$ given $\mathbf{R}_u^{(t)}$ is derived using the decoupled channel model (\ref{decouple_model}). For simplicity, the iteration index~$(t)$ is omitted in the following equations. The estimation process exploits the Markov chain $k_{u,m} \leftrightarrow \vecrho_{u,m,1:k_{u,m}} \leftrightarrow \vecx_{u,m} \leftrightarrow \vecr_{u,m}$. Using this Markov property, we can express the PME denoiser as
\begin{equation}
    \eta_u (\vecr_{u,m}) = \sum_{k=1}^{\uK_{u}} \Exop[\vecx_{u,m} | \vecr_{u,m}, k_{u,m}=k] \, p(k_{u,m}=k \mid \vecr_{u,m}). \label{eq:start_of_posterior_denoiser_b}
\end{equation}
Here, $p(k_{u,m}=k \mid \vecr_{u,m})$ is the posterior probability of the multiplicity $k_{u,m}$, and $\Exop[\vecx_{u,m} \mid \vecr_{u,m}, k_{u,m}=k]$ is the conditional mean. For simplicity, let $k_u$, $\vecrho_{u,1:k_u}$, $\vecr_u$, and $\vecx_u$ denote $k_{u,m}$, $\vecrho_{u,m,1:k_{u,m}}$, $\vecr_{u,m}$, and $\vecx_{u,m}$, respectively. Using Bayes' theorem, we can express the posterior probability~in~(\ref{eq:start_of_posterior_denoiser_b})~as
\begin{equation} \label{eqn:posterior_mults}
    p(k_{u}=k \mid \vecr_{u}) = \frac{p(\vecr_{u} \mid k_{u}=k) \, p(k_{u}=k)}{\sum_{l=0}^{\uK_{u}} p(\vecr_{u} \mid k_{u}=l) \, p(k_{u}=l)}.
\end{equation}
Using the Markov property and the prior on user positions, we can write the likelihood $p(\vecr_{u} \mid k_{u}=k)$ as
\begin{equation}
    p(\vecr_{u} \mid k_{u}=k) = \frac{1}{|\setD_u|^{k}}\int_{\setD_u^{k}}  p(\vecr_{u} \mid \vecrho_{u,1:k}) \, \mathrm{d}\vecrho_{u,1:k}. \label{eqn:likelihood_r_given_k}
\end{equation}    
It follows from (\ref{expression_x}) and (\ref{decouple_model}) that $\vecr_u$ is  distributed as 
$    
\sum_{i=1}^{k} \mathcal{C}\mathcal{N}(\veczero, \uN 
    \uP \matSigma(\rho_{u,i})) + \mathcal{C}\mathcal{N}(\veczero, \mathbf{T})
$, given $k_u=k$. The likelihood $p(\vecr_{u} \mid \vecrho_{u,1:k})$ in~(\ref{eqn:likelihood_r_given_k}) becomes
\begin{equation}   \label{eqn:explicit_likelihood_expression}
p(\vecr_{u} \mid \vecrho_{u,1:k}) = \textstyle \mathcal{C}\mathcal{N}(\vecr_{u}; \veczero, \mathbf{T} + \uN\uP 
\sum_{i=1}^{k} \matSigma(\rho_{u,i})).
\end{equation}
The conditional mean $\Exop[\vecx_{u} \mid \vecr_{u}, k_{u}=k]$ in~\eqref{eq:start_of_posterior_denoiser_b} is given by
\begin{align} 
    \Exop[\vecx_{u} \mid &\vecr_{u}, k_{u}=k] \notag \\
    & = \int_{\setD_u^{k}} \Exop[\vecx_{u} \mid \vecr_{u}, \vecrho_{u,1:k}] \, p(\vecrho_{u,1:k} \mid \vecr_{u})\, \mathrm{d}\vecrho_{u,1:k}. \label{eqn-bayesian-pme-17}
\end{align}
Note that~\cite[Sec.~12.5]{mmse_est}
\begin{align} \label{eqn:mmse_est}
    &\Exop[\vecx_{u} \mid \vecr_{u}, \vecrho_{u,1:k}] \notag \\
    &= \textstyle  ( 
    \sqrt{\uN \uP} 
    \sum_{i=1}^{k} \matSigma(\rho_{u,i}) ) (\mathbf{T} + \uN
    \uP 
    \sum_{i=1}^{k} \matSigma(\rho_{u,i}))^{-1} \vecr_{u} .
\end{align}
The posterior of the user positions in (\ref{eqn-bayesian-pme-17}) can be expressed as
\begin{align} 
    p(\vecrho_{u,1:k} \mid \vecr_{u})  = \frac{p(\vecr_{u} \mid \vecrho_{u,1:k})}{ \int_{\setD_u^{k}} p(\vecr_{u} \mid \vecrho'_{u,1:k}) \, \mathrm{d}\vecrho'_{u,1:k}}. \label{eqn-bayesian-pme-19}
\end{align}
We detail the derivations of \eqref{eq:start_of_posterior_denoiser_b}-\eqref{eqn-bayesian-pme-19} in Appendix~\ref{app:denoiser}. 

\subsubsection{Onsager Correction} \label{Subsec:onsager}
The Onsager correction ensures the convergence of the AMP algorithm by compensating for the correlations introduced during iterations. In multisource AMP~\cite{cakmak_2025_journal}, the Onsager term $\mathbf{Q}_u^{(t)} \in \mathbb{C}^{\uF \times \uF}$ in (\ref{amp_decoder_row3}) is given by
\begin{align}
[\mathbf{Q}_u^{(t)}]_{a,b} = \frac{1}{{\uM}} \sum_{m=1}^{{\uM}} \frac{\partial [\eta_{u,t}(\vecr_{u,m}^{(t)})]_b}{\partial [\vecr_{u,m}^{(t)}]_a}. \label{onsager_expression}
\end{align}
The derivation of this term is provided in Appendix~\ref{app:onsager}.

\subsubsection{Multiplicity Estimation} \label{subsec:type_est} Finally, we compute the posteriors $p(k_{u,m}=k\mid \vecr_{u,m})$ in~(\ref{eqn:posterior_mults}) and perform maximum a posteriori decoding as
\begin{align} \label{type_estimation_mapml} 
\widehat{k}_{u,m}  
&= \argmax_{k \in \{0,1,\dots,\uK_{u}\}} p(k_{u,m}=k\mid \vecr_{u,m}). 
\end{align} 
Then, we estimate the type $\widehat{\vect}$ as outlined in Section~\ref{Sec:SystemModel-Decoding}.

\subsection{Approximation Methods for Efficient Implementation}
The PME denoiser involves high-dimensional integrals in (\ref{eqn:likelihood_r_given_k}), (\ref{eqn-bayesian-pme-17}), and (\ref{eqn-bayesian-pme-19}) that are computationally prohibitive for high multiplicities. Therefore, we seek an efficient approximation. One could discretize the coverage area using a uniform discrete grid. However, the complexity of this approach grows exponentially with the number of points and the multiplicity, making it impractical for large-scale systems. Instead, we adopt Monte Carlo (MC) sampling. Specifically, we draw user positions independently from the uniform prior over $\setD_u$. Let $\{\vecrho^i_{u,1:k}\}_{i=1}^{\uN_{\mathrm{s}}}$ denote the MC samples of the positions of $k$ users. Using these samples, we approximate (\ref{eqn-bayesian-pme-17}) as 
\begin{equation}
    \Exop[\vecx_u \mid \vecr_u, k_u=k] \approx \frac{\sum_{i=1}^{\uN_{\mathrm{s}}} \Exop[\vecx_u \mid \vecr_u, \vecrho^i_{u,1:k}] p(\vecr_u \mid \vecrho^i_{u,1:k})}{\sum_{i=1}^{\uN_{\mathrm{s}}} p(\vecr_u \mid \vecrho^i_{u,1:k})}, \label{eqn-MCsampling1}
\end{equation}
and (\ref{eqn:posterior_mults}) as 
\begin{equation}
    p(k_u=k \mid \vecr_u) \approx \frac{\sum_{i=1}^{\uN_{\mathrm{s}}} p(\vecr_u \mid \vecrho^i_{u,1:k}) p(k_u=k)}{\sum_{i=1}^{\uN_{\mathrm{s}}} \sum_{l=0}^{\uK_{u}} p(\vecr_u \mid \vecrho^i_{u,1:l}) p(k_u=l)}. \label{eqn-MCsampling2}
\end{equation}
The details are provided in Appendix~\ref{app:approximation}. We summarize the proposed centralized decoder with MC sampling-based approximation in Algorithm~\ref{CentralizedDecoder}.

\begin{algorithm}[t!]
\caption{Centralized Decoder with Monte Carlo Sampling}
\textbf{Inputs:} Received signal $\mathbf{Y}$, codebook $\mathbf{C}$, factor $\uN\uP$, sampled positions $\{\vecrho^i_{u,1:k}\}_{i=1}^{\uN_{\mathrm{s}}}$ for $k \in [\uK_{u}]$, $u \in [\uU]$ \\
\textbf{Output:} Estimated type vector $\widehat{\vect}$ \\
\textbf{Initialization:} $\mathbf{Z}^{(0)} = \mathbf{Y}$, $\mathbf{X}_u^{(0)} = \veczero$, $u \in [\uU]$
\begin{algorithmic}[1]
    \Statex \textbf{1. AMP for Channel Estimation:}
    \State Precompute  $\{\sum_{j=1}^{k}\matSigma(\vecrho^i_{u,1:j})\}_{i=1}^{\uN_{\mathrm{s}}}$ for $k \in [\uK_{u}]$, $u \in [\uU]$
    \For{$t \gets 1$ to $\uT$}
        \For{$u \gets 1$ to $\uU$}
            \State $\mathbf{R}_u^{(t)} \gets \mathbf{C}_u^{\uH} \mathbf{Z}^{(t-1)} + \sqrt{\uN\uP} \mathbf{X}_u^{(t-1)}$
            \State Compute $\mathbf{T}^{(t)}$ as in (\ref{residual_expression})
            \State $\mathbf{X}_u^{(t)} \gets \eta_{u,t}(\mathbf{R}_u^{(t)})$ using \eqref{eq:start_of_posterior_denoiser_b}-\eqref{eqn-bayesian-pme-19}, (\ref{eqn-MCsampling1}) and (\ref{eqn-MCsampling2})
            \State Compute $\mathbf{Q}_u^{(t)}$ as in (\ref{onsager_expression}) and Appendix~\ref{app:onsager}
            \State $\mathbf{\Gamma}_u^{(t)} \gets \mathbf{C}_u \mathbf{X}_u^{(t)} - \frac{{\uM}}{\uN} \mathbf{Z}^{(t-1)} \mathbf{Q}_u^{(t)}$
        \EndFor
        \State $\mathbf{Z}^{(t)} \gets \mathbf{Y} - \sqrt{\uN\uP} \sum_{u=1}^{\uU} \mathbf{\Gamma}_u^{(t)}$
    \EndFor
    \Statex \textbf{2. Type Estimation:}
    \State Estimate $\{\widehat{\veck}_{u}\}_{u=1}^\uU$ as in (\ref{type_estimation_mapml}) using $p(k_{u,m}=k \mid \vecr_{u,m})$ computed in line 6 with \eqref{eqn-MCsampling2}
    \State $\widehat{\vect} \gets \sum_{u=1}^\uU \widehat{\veck}_u / \lVert \sum_{u=1}^\uU \widehat{\veck}_u\rVert_1$
\end{algorithmic}
\label{CentralizedDecoder}
\end{algorithm}

\subsection{Distributed Decoder}
To address the scalability in CF systems~\cite{cf_mimo_book}, we propose a distributed decoder inspired by dAMP~\cite{bai_larsson_damp}. Each AP locally processes its received signal and transmits likelihoods for each zone and each codeword to the CPU. Then, the likelihoods are aggregated as 
\begin{equation} p(\vecr_{u,m} \mid \vecrho_{u,1:k}) = \prod_{b=1}^{\uB} p_b(\vecr_{b,u,m} \mid \vecrho_{u,1:k}), \end{equation} where $p_b(\vecr_{b,u,m} \mid \vecrho_{u,1:k})$ is the local likelihood computed at AP $b$ with MC sampling as in (\ref{eqn-MCsampling2}). The posterior probability is then computed as 
\begin{align} 
p(k_{u,m} &= k \mid \vecr_{u,m}) \notag \\ &= \frac{p(k_{u,m}=k) \prod_{b=1}^{\uB} p_b(\vecr_{b,u,m} \mid \vecrho_{u,1:k})}{\sum_{l=0}^{\uK_{u}} p(k_{u,m}=l) \prod_{b=1}^{\uB} p_b(\vecr_{b,u,m} \mid \vecrho_{u,1:l})}, 
\end{align}
as detailed in Appendix~\ref{app:dist_AMP}. 
This design improves scalability by reducing the CPU's workload and fronthaul signaling through local processing at APs. The CPU then performs posterior computation and type estimation as in Section~\ref{SubSec-CentDec}.

\subsection{Complexity Analysis}
For notational convenience, we assume an equal number of active users per zone, i.e., $\uK_{u} = \uK_{\mathrm{a}} / \uU$, $u \in [\uU]$. Under this assumption, the complexity of the centralized decoder per AMP iteration is $O\big(\adjustedbar{\uM} \cdot (\uN + \uN_{\mathrm{s}} \cdot \uK_{u} \cdot \uF)\big)$, mainly due to matrix-vector multiplications and MC sampling in the denoiser. 
To reduce the cost, the decoder can exclude high multiplicities with negligible probabilities. Limiting the analysis to a maximum multiplicity $\uK_{\mathrm{max}}$ reduces the complexity to $O\big(\adjustedbar{\uM} \cdot (\uN + \uN_{\mathrm{s}} \cdot \uK_{\mathrm{max}} \cdot \uF)\big)$. 
For the distributed decoder, the complexity per AP is $O\big(\adjustedbar{\uM} \cdot (\uN + \uN_{\mathrm{s}} \cdot \uK_{\mathrm{max}} \cdot \uA)\big)$, while the CPU aggregates likelihoods and estimates type with complexity $O\big(\adjustedbar{\uM} \cdot \uB\big)$. Since $\uA \ll \uF$ in typical CF MIMO systems, the distributed approach significantly reduces computational cost.

\section{Simulation Results} \label{Sec:Sim}

We consider a CF massive MIMO system with a $3 \times 3$ square grid layout, where each zone is a nonoverlapping region as in Fig. \ref{fig:topology}. Each zone contains $\uK_u=20$ active users, transmitting $\uM_{\mathrm{a},u}=13$ active messages. %The set of active messages is the same across zones. 
With $\uU=9$, this results in $\uK_\mathrm{a}=180$ active users in total. As in \cite{gkiouzepi2024jointmessagedetectionchannel}, the LSFC is modeled as $\gamma_b(\rho) = 1/(1 + \left( |\rho - \nu_b|/d_0\right)^\alpha)$, where the pathloss exponent is $\alpha=3.67$ and the $\qty{3}{dB}$ cutoff distance is $d_0=\qty{13,57}{m}$. The side length of each zone is set to~$\qty{100}{m}$. Each AP, equipped with $\uA=4$ antennas, is evenly placed along the zone boundaries, with $\uB=40$ APs in total as shown in Fig. \ref{fig:topology}. As in \cite{gkiouzepi2024jointmessagedetectionchannel}, we define the received SNR as $\text{SNR}_{\text{rx}} = \text{SNR}_{\text{tx}} / (1 + (\varsigma/d_0)^\alpha)$, where $\varsigma$ is the distance between a zone centroid (green dot in Fig. \ref{fig:topology}) and its closest AP. The codewords $\{\vecc_{u,m}\}$ are independently drawn from a Gaussian random coding ensemble, $\vecc_{u,m} \sim_{\text{i.i.d.}} \mathcal{C}\mathcal{N}(0, 1/\uN)$.\footnote{While a Gaussian codebook is used for performance evaluation, the proposed decoders are compatible with arbitrary codebooks.} We fix the average codeword energy as $\uN \uP = 1$. We set the number of MC samples to $\uN_{\mathrm{s}} = 500$ and the number of AMP iterations to $\uT=20$. In each simulation, the set of active messages is drawn uniformly at random, their multiplicities are sampled from a multinomial distribution as described in Section~\ref{subsec:prior}, and a new codebook is generated. The TV distance~(\ref{eq:tv_dist_expression}) is averaged over $1000$ independent simulations.\footnote{The code to reproduce the numerical results is available at \url{https://github.com/okumuskaan/tuma_fading_cf}.}

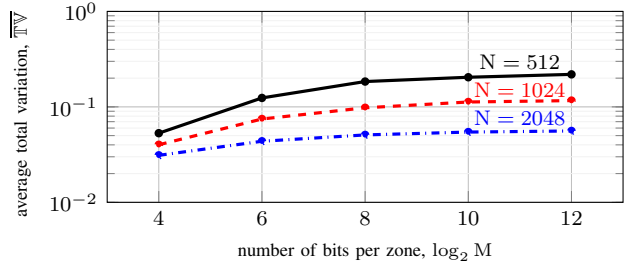
\begin{figure}[t!]
    \centering
    \begin{tikzpicture}
        \tikzstyle{every node}=[font=\footnotesize] 
        \begin{semilogyaxis}[
            scale only axis,
            width=2.7in,
            height=1.0in,
            grid=both,
            grid style={line width=.1pt, draw=gray!10},
            major grid style={line width=.2pt,draw=gray!50},
            %scale only axis,
            xmin=3,
            xmax=13,
            %extra x ticks={5},
            xlabel={\scriptsize number of bits per zone, $\log_2 \uM$},
            ymin=1e-2,
            ymax=1e0,
            ytick = {1e-4,1e-3,1e-2,1e-1, 1e0},
            ylabel={\scriptsize average total variation, $\dTV$},
            ylabel style = {xshift=-1mm},
            % legend pos=south east,
            legend style={draw=none,opacity=.9,at={(0.98,0.01)},anchor=south east},
        ]
        
       \addplot [dashed, color=red, mark=*, line width=1.2pt, mark size=1pt]
        table[row sep=crcr]{%
            4   0.04007872572783763 \\
            6   0.07464414219835254 \\
            8   0.09784519439837083 \\
            10  0.11248275862068964 \\
            12  0.11634555322221134 \\
        };
        
        \addplot [color=black, mark=*, line width=1.2pt, mark size=1pt]
        table[row sep=crcr]{%
            4    0.05302078548594286 \\ 
            6    0.12406371681818008 \\ 
            8    0.18425550005870353 \\ 
            10   0.20437333333333332 \\
            12   0.219023409432 \\
        };

        \addplot [dash dot, color=blue, mark=*, line width=1.2pt, mark size=1pt]
        table[row sep=crcr]{%
            4   0.03107116398821238 \\
            6   0.043620787839364523 \\
            8   0.05084375 \\
            10  0.054473333333333346 \\
            12  0.0559356355634 \\
        };
        \node[align = center,color=blue] at (axis cs:11,0.076) () {$\uN = 2048$}; % or: 0.065
        \node[align = center,color=red] at (axis cs:11,0.152) () {$\uN = 1024$};
        \node[align = center,color=black] at (axis cs:11,0.3) () {$\uN = 512$};
        
      \end{semilogyaxis}

    \end{tikzpicture}
    \vspace{-0.3cm}
    \caption{The average total variation $\dTV$ vs. the number of bits per zone $\log_2 \uM$ for $\text{SNR}_{\text{rx}} = \qty{-30}{dB}$ with centralized decoder.}
    \label{fig:tv_vs_mu}
    \vspace{-.3cm}
\end{figure}

In Fig.~\ref{fig:tv_vs_mu}, we show the average TV distance versus $\log_2 {\uM}$ for the centralized decoder for $\text{SNR}_{\text{rx}} = \qty{-30}{dB}$. Smaller ${\uM}$ leads to lower $\dTV$ because fewer codewords make estimation easier, despite higher message collisions. In contrast, larger ${\uM}$ makes estimation more challenging, but increasing the blocklength significantly improves performance.

\begin{figure}[t!]
    \centering
    \begin{tikzpicture}
        \tikzstyle{every node}=[font=\footnotesize] 
        \begin{semilogyaxis}[
        scale only axis,
        width=2.7in,
        height=1.0in,
        grid=both,
        grid style={line width=.1pt, draw=gray!10},
        major grid style={line width=.2pt,draw=gray!50},
    %scale only axis,
        xmin=-45,
        xmax=15,
        %extra x ticks={5},
        xlabel={\scriptsize received signal to noise ratio, $\text{SNR}_{\text{rx}}$ (dB)},
        ymin=1e-3,
        ymax=1,
        ytick = {1e-4,1e-3,1e-2,1e-1,1e0},
        ylabel={\scriptsize average total variation, $\dTV$},
        ylabel style = {xshift=-1mm},
         % legend pos=south east,
        legend style={draw=none,opacity=.9,at={(0.40,0.01)},anchor=south east},
        ]

        \addplot [color=black, mark=*, line width=1.2pt, mark size=1pt]
        table[row sep=crcr]{%
            -40  0.6380230389049589 \\ %0.44555026519999996 \\
            -30  0.09784519439837083 \\ %0.07592772489999998 \\
            -20  0.026257455 \\ %0.021442210499999993 \\
            -10  0.00984919 \\ %0.008936562300000011 \\
            0    0.00543319 \\ %0.00382449345434742 \\
            10   0.00369809 \\ %0.0027859043999999998\\
        };
        \addlegendentry{centralized};

        \addplot [color=red, mark=*, dashed, line width=1.2pt, mark size=1pt]
        table[row sep=crcr]{%
            -40  0.64167571 \\
            -30  0.11653672 \\
            -20  0.04962973 \\ % 0.10442210499999993 \\
            -10  0.03569764 \\ %0.085936562300000011 \\
            0    0.03446294 \\ %0.07382449345434742 \\
            10   0.03444081 \\ %0.0727859043999999998\\
        };
        \addlegendentry{distributed};
      \end{semilogyaxis}
    \end{tikzpicture}    
    \vspace{-0.3cm}
    \caption{The average total variation $\dTV$ vs. received signal to noise ratio $\text{SNR}_{\text{rx}}$ for $\uN=1024$ and ${\uM} = 2^8$.}
    \label{fig:tv_vs_snr}
    \vspace{-.3cm}
\end{figure}
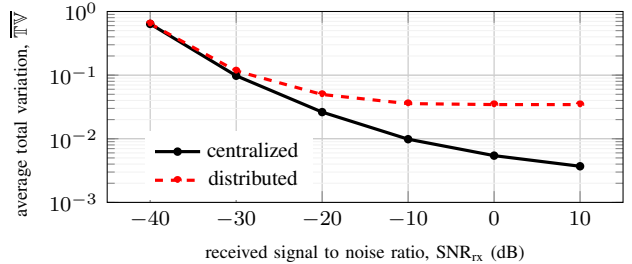

In Fig.~\ref{fig:tv_vs_snr}, we compare $\dTV$ for centralized and distributed decoders as a function of $\text{SNR}_{\text{rx}}$ for $\uN = 1024$ and ${\uM} = 2^8$. The centralized decoder consistently outperforms the distributed decoder. However, the distributed decoder reduces the computational cost at the CPU as well as the fronthaul rate. This makes it suitable for large-scale CF systems.

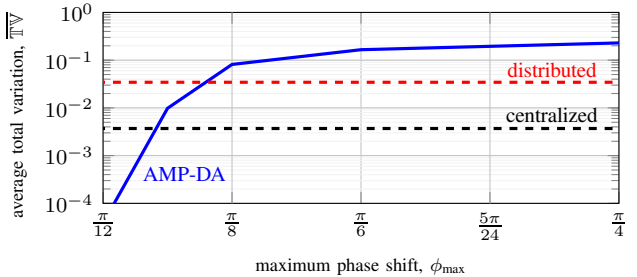
\begin{figure}[t!]
    \centering
    \begin{tikzpicture}
        \tikzstyle{every node}=[font=\footnotesize] 
        \begin{semilogyaxis}[
            scale only axis,
            width=2.7in,
            height=1.0in,
            grid=both,
            grid style={line width=.1pt, draw=gray!10},
            major grid style={line width=.2pt,draw=gray!50},
            xlabel={\scriptsize maximum phase shift, $\phi_{\text{max}}$},
            ymin=1e-4,
            ymax=1e0,
            xmin=0.2617993877991494,
            xmax=0.7853981633974483,
            ytick={1e-4, 1e-3, 1e-2, 1e-1, 1e0}, % Numerical positions for ticks
            xtick={0, 0.7853981633974483/3, 2*0.7853981633974483/3, 0.7853981633974483, 1.0471975511965976}, % Numerical positions for ticks
            xticklabels={$0$, $\frac{\pi}{12}$, $\frac{\pi}{6}$, $\frac{\pi}{4}$, $\frac{\pi}{3}$}, % LaTeX math labels
            extra x ticks={(1.0471975511965976/4 + 1.0471975511965976/2)/2, (1.0471975511965976/4 + 1.0471975511965976)/2, (0.7853981633974483+ 1.0471975511965976)/2}, % Additional ticks for grid lines
            extra x tick labels={$\frac{\pi}{8}$, $\frac{5\pi}{24}$, $\frac{7\pi}{12}$}, % No labels for extra ticks
            ylabel={\scriptsize average total variation, $\dTV$},
            %yminorticks=true,
            ylabel style = {xshift=-1mm},
        ]
            \addplot[color=blue, line width = 1.2pt]
            table[row sep=crcr]{
                0.2617 3.80470641e-05 \\ %1e-11 \\
                0.32724923 9.81863574e-03 \\
                0.3927 8.13835971e-02 \\ % 0.0042 \\
                0.5235 1.65600873e-01 \\ % 0.06493 \\
                %0.6545  \\ % 0.15275 \\
                0.7853 2.29710098e-01 \\ % 0.23102 \\
                %0.916298 2.94262255e-01 \\ % 0.2799 \\
                1.0471 2.94262255e-01 \\ % 0.32480 \\
            };
            \addplot[color=black, dashed, line width = 1.2pt]
            table[row sep=crcr]{
                -1 0.00369809 \\ % 0.0027859043999999998 \\
                10 0.00369809 \\ %0.0027859043999999998 \\
            };
            
            \addplot[color=red, dashed, line width = 1.2pt]
            table[row sep=crcr]{
                -1 0.03444081 \\ %0.0727859043999999998 \\
                10 0.03444081 \\ %0.0727859043999999998 \\
            };
            \node[align = center,color=red] at (axis cs:0.718,6.1e-2) () {distributed};
            \node[align = center,color=black] at (axis cs:0.718,7.0e-3) () {centralized};
            \node[align = center,color=blue] at (axis cs:0.3445, 4.2e-4) () {AMP-DA};
        \end{semilogyaxis}
    \end{tikzpicture}%
    \vspace{-0.3cm}
    \caption{The average total variation $\dTV$ vs. maximum phase shift $\phi_{\text{max}}$ for imperfect CSI with ${\uM}=2^8$ and $\text{SNR}_{\text{rx}} = \qty{10}{dB}$.}
    \label{fig:tv_vs_csi}
    \vspace{-.5cm}
\end{figure}

We next compare our decoders with AMP-DA \cite{qiao_gunduz_fl}, which lets the users pre-equalize the channel to obtain an effective AWGN channel model, and then apply scalar AMP~\cite[Sec.~IV-C]{Meng_NOMA_2021} for type estimation. We emphasize that the pre-equalization step relies on the availability of CSI at the users. Here, we assume that each user has an imperfect knowledge $\widehat{\vech}$ of its channel vector $\vech$. Specifically, $\widehat{\vech} = \vech \odot e^{j\vecphi}$ with $\vecphi \sim_{\text{i.i.d.}} \Unif(0, \phi_{\text{max}})$ is used to perform pre-equalization. In contrast, our decoders do not use CSI, neither at the users nor the receiver, and are thus insensitive to $\phi_{\text{max}}$. In Fig.~\ref{fig:tv_vs_csi}, we demonstrate that our centralized and distributed decoders outperform AMP-DA when the maximum phase shift $\phi_{\text{max}}$ exceeds approximately $\pi/10$ and $\pi/9$, respectively.

\section{Conclusion}
We extended the TUMA framework proposed in~\cite{ngo2024_tuma} to fading channels. Specifically, we proposed centralized and distributed decoders for TUMA over CF massive MIMO systems. The centralized decoder demonstrates superior performance and robustness, particularly in the high SNR regime. The distributed decoder, while less accurate, provides a cost-effective solution for large-scale systems. As in~\cite{gkiouzepi2024jointmessagedetectionchannel}, our decoders rely on multisource AMP, suitably modified to handle message collisions, and use a location-based codeword partition to mitigate such collisions.

%\section*{Acknowledgment}
%This work was supported in part by the Swedish Research Council under grants 2021-04970 and 2022-04471, and by the Swedish Foundation for Strategic Research.

\appendix

\subsection{Derivations for the Denoiser} \label{app:denoiser}
For notational simplicity, as in Section~\ref{subsec:denoiser}, we omit the AMP iteration index $(t)$ and let $k_u$, $\vecrho_{u,1:k_u}$, $\vecr_u$, and $\vecx_u$ denote $k_{u,m}$, $\vecrho_{u,m,1:k_{u,m}}$, $\vecr_{u,m}$, and $\vecx_{u,m}$, respectively. These simplifications are valid since the derivation structure is identical across iterations and codewords. We also rely on the Markov chain $k_{u} \leftrightarrow \vecrho_{u,1:k_{u}} \leftrightarrow \vecr_{u}$.

\subsubsection{Posterior Probability of Multiplicities} \label{app:posterior_prob}
The posterior probability $p(k_u =k \mid \vecr_u)$ in~\eqref{eqn:posterior_mults} is derived via Bayes' theorem. The likelihood $p(\vecr_u \mid k_u=k)$ in~\eqref{eqn:likelihood_r_given_k} follows~as 

\vspace{-15pt}
\begin{subequations}
    \begin{align}
    &p(\vecr_u \mid k_u=k) \notag \\
    &=\int_{\setD_u^{k}} p(\vecr_u \mid \vecrho_{u,1:k}, k_u=k) p(\vecrho_{u,1:k} \mid k_u=k)\, \mathrm{d}\vecrho_{u,1:k} \label{eqn-deriv-p_ruku_step1}\\
    &=\frac{1}{|\setD_u|^{k}} \int_{\setD_u^{k}} p(\vecr_u \mid \vecrho_{u,1:k})\, \mathrm{d}\vecrho_{u,1:k}, \label{eqn-deriv-p_ruku_step2}
    \end{align}
\end{subequations}
where in (\ref{eqn-deriv-p_ruku_step1}) we used the law of total probability, and in~(\ref{eqn-deriv-p_ruku_step2}) we applied the Markov chain properly and the assumption of uniform prior over positions $p(\vecrho_{u,1:k_u} \mid k_u=k)= 1/|\setD_u|^{k}$ (see Section~\ref{subsec:prior}). 

\subsubsection{Likelihood of Received Signal} \label{app:likelihood}
Based on the effective channel model (\ref{decouple_model}), the received signal $\vecr_u$ is distributed as the sum of independent complex Gaussian signals: $\sum_{i=1}^{k_u} \mathcal{CN}(\veczero, \uN\uP \matSigma(\rho_{u,i})) + \mathcal{CN}(\veczero, \mathbf{T})$, given $k_u=k$. The resulting likelihood $p(\vecr_u \mid \vecrho_{u,1:k})$ is given in~\eqref{eqn:explicit_likelihood_expression}.

\subsubsection{Posterior Distribution of Positions} \label{app:posterior_positions}
The posterior $p(\vecrho_{u,1:k} \mid \vecr_u)$ in~\eqref{eqn-bayesian-pme-19} is derived as
\begin{subequations} \label{app:channel_estimator_third}
\begin{align}
p&(\vecrho_{u, 1:k} \mid \vecr_u) =p(\vecrho_{u,1:k_u} \mid \vecr_u, k_u=k) \notag \\ 
&= \frac{p(\vecr_u\mid \vecrho_{u, 1:k_u},k_u=k)p(\vecrho_{u,1:k_u}\mid k_u=k)}{p(\vecr_u\mid k_u=k)} \label{eqn-deriv-posteriorpositions_step1} \\
&= \frac{p(\vecr_u\mid \vecrho_{u,1:k})p(\vecrho_{u,1:k}\mid k_u=k)}{\int_{\setD_u^{k}}p(\vecr_u\mid \vecrho'_{u,1:k}) p(\vecrho'_{u,1:k}\mid k_u=k) \, \mathrm{d}\vecrho'_{u,1:k}} \label{eqn-deriv-posteriorpositions_step2} \\
&= \frac{p(\vecr_u\mid \vecrho_{u,1:k})/|\setD_u|^{k}}{\int_{\setD_u^{k}}(p(\vecr_u\mid \vecrho'_{u,1:k}) /|\setD_u|^{k}) \, \mathrm{d}\vecrho'_{u,1:k}} \label{eqn-deriv-posteriorpositions_step3} \\
&= \frac{p(\vecr_u\mid \vecrho_{u,1:k})}{\int_{\setD_u^{k}}p(\vecr_u\mid \vecrho'_{u,1:k}) \, \mathrm{d}\vecrho'_{u,1:k}}. \label{eqn-deriv-posteriorpositions_step4}
\end{align}
\end{subequations}
Here, (\ref{eqn-deriv-posteriorpositions_step1}) follows from the Bayes' theorem, (\ref{eqn-deriv-posteriorpositions_step2}) from the law of total probability and the Markov chain, (\ref{eqn-deriv-posteriorpositions_step3}) from the uniform prior $p(\vecrho_{u,1:k_u} \mid k_u=k)=1/|\setD_u|^k$.

\subsubsection{Conditional Mean of \texorpdfstring{$\vecx_u$}{x\_u}} \label{app:conditional_mean}
Using the law of total probability, we express the conditional mean $\Exop[\vecx_u \mid \vecr_u, k_u=k]$ as in~\eqref{eqn-bayesian-pme-17}. The inner term $\Exop[\vecx_u \mid \vecr_u, \vecrho_{u,1:k}]$ corresponds to the minimum mean square error estimator with complex Gaussian priors, given in~\eqref{eqn:mmse_est}.

Finally, the denoiser $\eta(\vecr_u)$, expanded as in \eqref{eq:start_of_posterior_denoiser_b}, is computed by combining \eqref{eqn:posterior_mults}–\eqref{eqn:mmse_est}.

\subsection{Derivations for the Onsager Correction} \label{app:onsager}

As in Appendix~\ref{app:denoiser}, we omit the iteration index~$(t)$ and let $\vecrho_{u,1:k_u}$ and $\vecr_{u}$ denote $\vecrho_{u,m,1:k_{u,m}}$ and $\vecr_{u,m}$, respectively. The Onsager correction term $\mathbf{Q}_u \in \mathbb{C}^{\uF \times \uF}$ is defined in~\eqref{onsager_expression} as the average Jacobian matrix of the denoiser function $\eta(\cdot)$. Here, the derivatives are found using Wirtinger derivatives for complex variables, where $\partial (\cdot)/\partial r = \left( \partial (\cdot)/\partial r_x - j \partial (\cdot)/\partial r_y   \right)/2$, with $r = r_x + jr_y$~\cite[App.~A]{wirtinger}. Combining \eqref{eqn:posterior_mults}–\eqref{eqn:mmse_est}, we decompose the denoiser component $[\eta(\vecr_u)]_b$~as
\begin{align} 
[\eta(\vecr_u)]_b &= [\vecr_u]_b \underbrace{\sum_{k=1}^{\uK_u} \underbrace{\frac{A_b(\vecr_u,k)}{B(\vecr_u,k)}}_{F_b(\vecr_u, k)} \cdot \underbrace{\frac{C(\vecr_u,k)}{D(\vecr_u)}}_{G(\vecr_u, k)}}_{H_b(\vecr_u)},
\end{align} 
where 
\begin{subequations}    
\begin{align} A_b(\vecr_u,k) &= \int_{\setD_u^k} c_{b,\vecrho_{u,1:k}} \, p(\vecr_u \mid  \vecrho_{u,1:k}) \, \mathrm{d}\vecrho_{u,1:k}, \\
c_{b,\vecrho_{u,1:k}} &= \frac{\sqrt{\uN\uP} [\textstyle \sum_{i=1}^k\matSigma(\rho_{u,i})]_{b,b}}{[\mathbf{T}]_{b,b} + \uN\uP [\sum_{i=1}^k\matSigma(\rho_{u,i})]_{b,b} }, \label{deriv-eqn-cbp}  \\
B(\vecr_u,k) &= \int_{\setD_u^k} p(\vecr_u \mid  \vecrho_{u,1:k}) \, \mathrm{d}\vecrho_{u,1:k}, \\ 
C(\vecr_u,k) &= \int_{\setD_u^k} p(\vecr_u \mid  \vecrho_{u,1:k}) \, p(k_u = k) \, \mathrm{d}\vecrho_{u,1:k}, \\ D(\vecr_u) &= \sum_{l=0}^{\uK_u} \int_{\setD_u^k} p(\vecr_u \mid  \vecrho_{u,1:l}) \, p(k_u = l) \, \mathrm{d}\vecrho_{u,1:l} . \end{align}
\end{subequations}
Note that (\ref{deriv-eqn-cbp}) is derived using the fact that $\mathbf{T}$ and $\sum_{i=1}^k \matSigma(\rho_{u,i})$ are both diagonal matrices. Then, the derivative of the denoiser is computed as 
\begin{subequations}
    \begin{align} \frac{\partial [\eta(\vecr_u)]_b}{\partial [\vecr_u]_a} &= \frac{\partial}{\partial [\vecr_u]_a} \left( [\vecr_u]_b H_b(\vecr_u) \right) \\ &= \delta(a-b) H_b(\vecr_u) + [\vecr_u]_b \frac{\partial H_b(\vecr_u)}{\partial [\vecr_u]_a}, 
    \end{align} \label{eqn_deriv30}
\end{subequations}
where we note that $H_b(\vecr_u) \in \mathbb{R}$. Expanding $H_b(\vecr_u)$, we have
\begin{equation}
    \frac{\partial H_b(\vecr_u)}{\partial [\vecr_u]_a} = G(\vecr_u, k) \frac{\partial F_b(\vecr_u, k)}{\partial [\vecr_u]_a} + F_b(\vecr_u, k) \frac{\partial G(\vecr_u,k)}{\partial [\vecr_u]_a}. \label{eqn_deriv31}
\end{equation}
The derivatives of $F_b(\vecr_u,k)$ and $G(\vecr_u,k)$ are given by
\begin{subequations} \label{eqn_deriv32}
    \begin{align} \frac{\partial F_b(\vecr_u,k)}{\partial [\vecr_u]_a} &= \frac{\frac{\partial A_b(\vecr_u,k)}{\partial [\vecr_u]_a} - F_b(\vecr_u,k) \frac{\partial B(\vecr_u,k)}{\partial [\vecr_u]_a}}{B(\vecr_u,k)}, \label{eqn_deriv_Fb}\\ \frac{\partial G(\vecr_u,k)}{\partial [\vecr_u]_a} &= \frac{\frac{\partial C(\vecr_u,k)}{\partial [\vecr_u]_a} - G(\vecr_u,k) \frac{\partial D(\vecr_u)}{\partial [\vecr_u]_a}}{D(\vecr_u)}. \label{eqn_deriv_G}
    \end{align}
\end{subequations}
The derivatives on the right hand sides of \eqref{eqn_deriv_Fb} and \eqref{eqn_deriv_G} are
\begin{subequations} \label{deriv_eqn_allcomps}
    \begin{align} \frac{\partial A_b(\vecr_u,k)}{\partial [\vecr_u]_a} &= \int_{\setD_u^k} c_{b,\vecrho_{u,1:k}} \frac{\partial p(\vecr_u \mid \vecrho_{u,1:k})}{\partial [\vecr_u]_a} \, \mathrm{d}\vecrho_{u,1:k}, \\ 
    \frac{\partial B(\vecr_u,k)}{\partial [\vecr_u]_a} &= \int_{\setD_u^k} \frac{\partial p(\vecr_u \mid \vecrho_{u,1:k})}{\partial [\vecr_u]_a} \, \mathrm{d}\vecrho_{u,1:k}, \\
    \frac{\partial C(\vecr_u,k)}{\partial [\vecr_u]_a} &= \int_{\setD_u^k} p(k_u=k)\, \frac{\partial p(\vecr_u \mid \vecrho_{u,1:k})}{\partial [\vecr_u]_a} \, \mathrm{d}\vecrho_{u,1:k}, \\
    \frac{\partial D(\vecr_u)}{\partial [\vecr_u]_a} &= \sum_{l=0}^{\uK_u} \int_{\setD_u^k} p(k_u=l) \, \frac{\partial p(\vecr_u \mid \vecrho_{u,1:l})}{\partial [\vecr_u]_a} \, \mathrm{d}\vecrho_{u,1:l}. \end{align}
\end{subequations}
Since $p(\vecr_u \mid \vecrho_{u,1:k}) = \mathcal{C}\mathcal{N}(\vecr_u; \veczero, \mathbf{T}+ \uN \uP \sum_{i=1}^k \matSigma(\rho_{u,i}))$, its derivative is
\begin{align}
    &\frac{\partial p(\vecr_u \mid \vecrho_{u,1:k})}{\partial [\vecr_u]_a} \notag \\
    &= p(\vecr_u \mid \vecrho_{u,1:k}) \frac{\partial}{\partial [\vecr_u]_a} \sum_{f=1}^{\uF} \frac{-([\vecr_u]^2_{f,x} + [\vecr_u]^2_{f,y})}{[\mathbf{T}]_{f,f} + \uN \uP [\sum_{i=1}^k \matSigma(\rho_{u,i})]_{f,f}}, \label{eqn_deriv_expression}
\end{align}
with $[\vecr_u]_a = [\vecr_u]_{a,x} + j [\vecr_u]_{a,y}$. Note that
\begin{align}
    &\frac{\partial \sum_{f=1}^{\uF} \frac{-([\vecr_u]^2_{f,x} + [\vecr_u]^2_{f,y})}{[\mathbf{T} + \uN \uP \sum_{i=1}^k \matSigma(\rho_{u,i})]_{f,f}} }{\partial [\vecr_u]_{a,x}}  \notag \\ &\qquad \qquad \qquad \qquad = \frac{-2[\vecr_u]_{a,x}}{[\mathbf{T} + \uN \uP \sum_{i=1}^k \matSigma(\rho_{u,i})]_{a,a}}, \\
    &\frac{\partial \sum_{f=1}^{\uF} \frac{-([\vecr_u]^2_{f,x} + [\vecr_u]^2_{f,y})}{[\mathbf{T} + \uN \uP \sum_{i=1}^k \matSigma(\rho_{u,i})]_{f,f}} }{\partial [\vecr_u]_{a,y}} \notag \\ &\qquad \qquad \qquad \qquad = \frac{-2[\vecr_u]_{a,y}}{[\mathbf{T} + \uN \uP \sum_{i=1}^k \matSigma(\rho_{u,i})]_{a,a}}.
\end{align} 
By applying Wirtinger derivative, we obtain
\begin{align}
    &\frac{\partial \sum_{f=1}^{\uF} \frac{-([\vecr_u]^2_{f,x} + [\vecr_u]^2_{f,y})}{[\mathbf{T} + \uN \uP \sum_{i=1}^k \matSigma(\rho_{u,i})]_{f,f}} }{\partial [\vecr_u]_{a}} \notag \\
    & \qquad \qquad \qquad \qquad  = \frac{-[\vecr_u]^{\ast}_{a}}{[\mathbf{T} + \uN \uP \sum_{i=1}^k \matSigma(\rho_{u,i})]_{a,a}}. \label{eqn_deriv_wirting_first}
\end{align}
By using \eqref{eqn_deriv_wirting_first} in \eqref{eqn_deriv_expression}, we have
\begin{align}
    \frac{\partial p(\vecr_u \mid \vecrho_{u,1:k})}{\partial [\vecr_u]_a}
    &= \frac{-[\vecr_u]^{\ast}_{a} \, p(\vecr_u \mid \vecrho_{u,1:k})}{[\mathbf{T} + \uN \uP \sum_{i=1}^k \matSigma(\rho_{u,i})]_{a,a}}. \label{deriv_eq_main_simpler}
\end{align}
Substituting \eqref{deriv_eq_main_simpler} into \eqref{deriv_eqn_allcomps}, we obtain
\begin{subequations}    
\begin{align}
    &\frac{\partial A_b(\vecr_u,k)}{\partial [\vecr_u]_a} \notag \\
    & = -[\vecr_u]_a^\ast \int_{\setD_u^k} \frac{c_{b,\vecrho_{u,1:k}} p(\vecr_u \mid \vecrho_{u,1:k})}{[\mathbf{T} + \uN \uP \sum_{i=1}^k \matSigma(\rho_{u,i})]_{a,a}} \, \mathrm{d}\vecrho_{u,1:k}, \\
    &\frac{\partial B(\vecr_u,k)}{\partial [\vecr_u]_a} \notag \\
    & = -[\vecr_u]_a^\ast \int_{\setD_u^k} \frac{p(\vecr_u \mid \vecrho_{u, 1:k})}{[\mathbf{T} + \uN \uP \sum_{i=1}^k \matSigma(\rho_{u,i})]_{a,a}} \, \mathrm{d}\vecrho_{u,1:k}, \\
    &\frac{\partial C(\vecr_u,k)}{\partial [\vecr_u]_a} \notag \\
    & = -[\vecr_u]_a^\ast \int_{\setD_u^k} \frac{p(k_u=k)\,  p(\vecr_u \mid \vecrho_{u,1:k})}{[\mathbf{T} + \uN \uP \sum_{i=1}^k \matSigma(\rho_{u,i})]_{a,a}} \, \mathrm{d}\vecrho_{u,1:k}, \\
    &\frac{\partial D(\vecr_u)}{\partial [\vecr_u]_a} \notag \\
    & = -[\vecr_u]_a^\ast \sum_{l=0}^{\uK_u} \int_{\setD_u^l} \frac{p(k_u=l)\, p(\vecr_u \mid \vecrho_{u,1:l})}{[\mathbf{T} + \uN \uP \sum_{i=1}^k \matSigma(\rho_{u,i})]_{a,a}} \, \mathrm{d}\vecrho_{u,1:l}. 
\end{align} \label{deriv_local_explicit}
\end{subequations}
Finally, by combining \eqref{eqn_deriv30}, \eqref{eqn_deriv31}, \eqref{eqn_deriv32}, and \eqref{deriv_local_explicit}, we obtain the derivative $\partial [\eta(\vecr_u)]_b / \partial [\vecr_u]_a$.

\subsection{Derivations for Monte Carlo Sampling-Based Approximated Denoiser}\label{app:approximation}

Here, we derive the approximated expressions for the Bayesian PME and the posterior probability of multiplicities using MC sampling, as used in \eqref{eqn-MCsampling1} and \eqref{eqn-MCsampling2}. As in Appendix~\ref{app:denoiser}, we omit iteration index~$(t)$ and let $k_u$, $\vecrho_{u,1:k_u}$, $\vecr_u$, and $\vecx_u$ denote $k_{u,m}$, $\vecrho_{u,m,1:k_{u,m}}$, $\vecr_{u,m}$, and $\vecx_{u,m}$, respectively. Using MC sampling, $\vecrho_{u,1:k}$ is sampled independently from the uniform prior over $\setD_u$, for $k=1,\dots,\uK$. Let $\{\vecrho^i_{u,1:k}\}_{i=1}^{\uN_{\mathrm{s}}}$ denote these samples for $k_u=k$. We approximate the integral $\int_{\setD_u^{k}}f(\vecrho_{u,1:k})\mathrm{d}\vecrho_{u,1:k}$ as $\frac{|\setD_u|^{k}}{\uN_{\mathrm{s}}}\sum_{i=1}^{\uN_{\mathrm{s}}} f(\vecrho^i_{u,1:k})$. As a result, the PME is approximated as
\begin{subequations}    
\begin{align}
\Exop&[\vecx_u \mid \vecr_u, k_u=k]  \notag \\
&= \frac{\int_{\setD_u^{k}} \Exop[\vecx_u \mid \vecr_u, \vecrho_{u,1:k}]\,  p(\vecr_u \mid \vecrho_{u,1:k})  \, \mathrm{d}\vecrho_{u,1:k}}{\int_{\setD_u^{k}}  p(\vecr_u \mid \vecrho_{u,1:k})  \, \mathrm{d}\vecrho_{u,1:k}} \\
&\approx \frac{\frac{|\setD_u|^{k}}{\uN_{\mathrm{s}}}\sum_{i=1}^{\uN_{\mathrm{s}}} \Exop[\vecx_u \mid \vecr_u, \vecrho^i_{u,1:k}] \, p(\vecr_u \mid \vecrho^i_{u,1:k})}{\frac{|\setD_u|^{k}}{\uN_{\mathrm{s}}}\sum_{i=1}^{\uN_{\mathrm{s}}} p(\vecr_u \mid \vecrho^i_{u,1:k})}\\
&= \frac{\sum_{i=1}^{\uN_{\mathrm{s}}} \Exop[\vecx_u \mid \vecr_u, \vecrho^i_{u,1:k}] \, p(\vecr_u \mid \vecrho^i_{u,1:k})}{\sum_{i=1}^{\uN_{\mathrm{s}}} p(\vecr_u \mid \vecrho^i_{u,1:k})},
\end{align}
\end{subequations}
where $p(\vecr_u \mid \vecrho^i_{u,1:k})$ is the likelihood of the received signal given the sampled positions, computed as in~\eqref{eqn:explicit_likelihood_expression}. The posterior probability $p(k_u=k \mid \vecr_u)$ and the integrals in the Onsager term can be obtained using similar steps.

\subsection{Distributed Multisource AMP}\label{app:dist_AMP}

The distributed version of the multisource AMP algorithm decentralizes computations across APs.

\subsubsection{Distributed Likelihood Computations}
In the distributed AMP setup, each AP $b$ processes its local effective received signal $\vecr_{b,u,m} \in \mathbb{C}^{\uA}$ for each codeword $\vecc_{u,m}$. The local likelihood $p_b(\vecr_{b,u,m} \mid \vecrho_{u,1:k})$ is computed as
\begin{multline}
p_b(\vecr_{b,u,m} \mid \vecrho_{u,1:k}) \\ = \frac{1}{\pi^{\uA} \text{det}(\text{Cov}_b)} \exp \mathopen{}\left(-\vecr_{b,u,m}^{\uH} \text{Cov}_b^{-1} \vecr_{b,u,m}\right), \label{eqn_dist_likelihood}
\end{multline}
where $\text{Cov}_b$ is the local covariance matrix given by
$\text{Cov}_b = \mathbf{T}_b + \uN \uP \sum_{i=1}^k \matSigma_b(\rho_{u,i})$ and $\text{det}(\cdot)$ denotes the determinant. Here, $\mathbf{T}_b$ is the covariance matrix of the local residual noise computed at AP~$b$ and $\matSigma_b(\rho)=\gamma_b(\rho)\mathbf{I}_{\uA}$ is the LSFC between the user in position~$\rho$ and AP~$b$. Let $\text{Cov}$ denote the global covariance matrix given by $\text{Cov} = \mathbf{T} + \uN \uP \sum_{i=1}^k \matSigma(\rho_{u,i})$. The aggregation of the likelihood at the CPU is performed as
\begin{subequations}    
\begin{align}
&p(\vecr_{u,m} \mid \vecrho_{u,1:k}) \notag \\
&= \frac{1}{\pi^\uF \text{det}(\text{Cov})}\exp(-\vecr_{u,m}^\uH \text{Cov}^{-1} \vecr_{u,m})\\
&= \frac{1}{\prod_{b=1}^\uB \pi^\uA \text{det}(\text{Cov}_b)}\exp\left(- \sum_{b=1}^\uB \sum_{a=1}^\uA \frac{|[\vecr_{u,m}]_{a+b\uA}|^2}{[\text{Cov}_b]_{a,a}}  \right)\\
&= \prod_{b=1}^{\uB} p_b(\vecr_{b,u,m} \mid \vecrho_{u,1:k}).
\end{align}
\end{subequations}

\subsubsection{Posterior Probability in Distributed AMP}
Using the distributed likelihood \eqref{eqn_dist_likelihood}, the posterior probability $p(k_{u,m}=k \mid \vecr_{u,m})$ is computed as
\begin{align} 
&p(k_{u,m}=k \mid \vecr_{u,m}) \notag\\ & \qquad  = \frac{p(k_{u,m}=k) \prod_{b=1}^{\uB} p_b(\vecr_{b,u,m} \mid \vecrho_{u,1:k})}{\sum_{l=0}^{\uK_u} p(k_{u,m}=l) \prod_{b=1}^{\uB} p_b(\vecr_{b,u,m} \mid \vecrho_{u,1:l})}.
\end{align}

\subsubsection{Distributed Onsager Correction}
The Onsager correction term in distributed AMP is computed locally at each AP~$b$~as
\begin{equation}
[\mathbf{Q}_{u,b}]_{a,c} = \frac{1}{{\uM}} \sum_{m=1}^{{\uM}} \frac{\partial [\eta_{u}(\vecr_{b,u,m})]_c}{\partial [\vecr_{b,u,m}]_a},
\end{equation}
and aggregated at the CPU as
\begin{equation}
[\mathbf{Q}_u]_{a,c} = \sum_{b=1}^{\uB} [\mathbf{Q}_{u,b}]_{a,c}.
\end{equation}

\subsubsection{Distributed Likelihood Computations}
In the distributed AMP setup, each AP $b$ processes its local effective received signal $\vecr_{b,u,m} \in \mathbb{C}^{\uA}$ for each codeword $\vecc_{u,m}$. The local likelihood $p_b(\vecr_{b,u,m} \mid \vecrho_{u,1:k})$ is computed as
\begin{multline}
p_b(\vecr_{b,u,m} \mid \vecrho_{u,1:k})  \\ = \frac{1}{\pi^{\uA} \text{det}(\text{Cov}_b)} \exp \mathopen{}\left(-\vecr_{b,u,m}^{\uH} \text{Cov}_b^{-1} \vecr_{b,u,m} \right), \label{eqn_dist_likelihood_2}
\end{multline}
where $\text{Cov}_b$ is the local covariance matrix given by
\begin{equation}
\text{Cov}_b = \mathbf{T}_b + \uN \uP \sum_{i=1}^k \matSigma_b(\rho_{u,i}).
\end{equation}
Here, $\mathbf{T}_b$ is the covariance matrix of the local residual noise computed at AP~$b$ and $\matSigma_b(\rho)=\gamma_b(\rho)\mathbf{I}_{\uA}$ is the LSFC between the user in position~$\rho$ and AP~$b$. The aggregated likelihood at the CPU is then computed as
\begin{equation}
p(\vecr_{u,m} \mid \vecrho_{u,1:k}) = \prod_{b=1}^{\uB} p_b(\vecr_{b,u,m} \mid \vecrho_{u,1:k}).
\end{equation}

\subsubsection{Posterior Probability in Distributed AMP}
Using the distributed likelihood \eqref{eqn_dist_likelihood_2}, the posterior probability $p(k_{u,m} \mid \vecr_{u,m})$ is computed as
\begin{align} 
&p(k_{u,m}=k \mid \vecr_{u,m}) \notag\\ & \qquad  = \frac{p(k_{u,m}=k) \prod_{b=1}^{\uB} p_b(\vecr_{b,u,m} \mid \vecrho_{u,1:k})}{\sum_{l=0}^{\uK_{u}} p(k_{u,m}=l) \prod_{b=1}^{\uB} p_b(\vecr_{b,u,m} \mid \vecrho_{u,1:l})}.
\end{align}

\subsubsection{Distributed Onsager Correction}
The Onsager correction term in distributed AMP is computed locally at each AP $b$ as
\begin{equation}
[\mathbf{Q}_{u,b}]_{a,c} = \frac{1}{{\uM}} \sum_{m=1}^{{\uM}} \frac{\partial [\eta_{u,t}(\vecr_{b,u,m})]_c}{\partial [\vecr_{b,u,m}]_a},
\end{equation}
and aggregated at the CPU as
\begin{equation}
[\mathbf{Q}_u]_{a,c} = \sum_{b=1}^{\uB} [\mathbf{Q}_{u,b}]_{a,c}.
\end{equation}

%We are indebted to Michael Shell for maintaining and improving
%\texttt{IEEEtran.cls}. 

%%%%%%
%% To balance the columns at the last page of the paper use this
%% command:
%%
%\enlargethispage{-1.2cm} 
%%
%% If the balancing should occur in the middle of the references, use
%% the following trigger:
%%
%\IEEEtriggeratref{7}
%%
%% which triggers a \newpage (i.e., new column) just before the given
%% reference number. Note that you need to adapt this if you modify
%% the paper.  The "triggered" command can be changed if desired:
%%
%\IEEEtriggercmd{\enlargethispage{-20cm}}
%%
%%%%%%

%%%%%%
%% References:
%% We recommend the usage of BibTeX:
%%
%\bibliographystyle{IEEEtran}
%\bibliography{definitions,bibliofile}
%%
%% where we here have assumed the existence of the files
%% definitions.bib and bibliofile.bib.
%% BibTeX documentation can be obtained at:
%% http://www.ctan.org/tex-archive/biblio/bibtex/contrib/doc/
%%%%%%

\input{main.bbl}

\end{document}

%% file: figs/cf_topology.tex
\definecolor{darkgreen}{rgb}{0.0, 0.5, 0.0}
\definecolor{yelloworange}{rgb}{1.0, 0.65, 0.05}

\begin{tikzpicture}

% Add orange RUs and labels
\foreach \x/\y/\label in {
    0/0/RU 1, 1/0/RU 2, 2/0/RU 3, 3/0/RU 4,
    0/1/RU 5, 1/1/RU 6, 2/1/RU 7, 3/1/RU 8,
    0/2/RU 9, 1/2/RU 10, 2/2/RU 11, 3/2/RU 12,
    0/3/RU 13, 1/3/RU 14, 2/3/RU 15, 3/3/RU 16
} {
    \node[fill=yelloworange, rectangle, opacity=1, scale=1.5] at (\x*4, \y*4) {};
}

% Draw outer black border
\def\D{4} % Distance between APs
\draw[black, very thick] (0, 0) rectangle (3*\D, 3*\D);

% Draw inner gray zone boundaries (3x3 zones)
\foreach \x in {1,2} {
    \draw[gray, thick] (\x*\D, 0) -- (\x*\D, 3*\D);
}
\foreach \y in {1,2} {
    \draw[gray, thick] (0, \y*\D) -- (3*\D, \y*\D);
}

\foreach \x/\y/\label in {
    0/0/RU 1, 1/0/RU 2, 2/0/RU 3, 3/0/RU 4,
    0/1/RU 5, 1/1/RU 6, 2/1/RU 7, 3/1/RU 8,
    0/2/RU 9, 1/2/RU 10, 2/2/RU 11, 3/2/RU 12,
    0/3/RU 13, 1/3/RU 14, 2/3/RU 15, 3/3/RU 16
} {
    \node[fill=yelloworange, rectangle, opacity=1, scale=2] at (\x*4, \y*4) {};

    % Additional vertical points
    \ifnum\y<3 % Only add vertical points if there's a row below
        \node[fill=yelloworange, rectangle, opacity=1, scale=2] at (\x*4, \y*4 + 2) {};
    \fi
    
    % Additional horizontal points
    \ifnum\x<3 % Only add horizontal points if there's a column to the right
        \node[fill=yelloworange, rectangle, opacity=1, scale=2] at (\x*4 + 2, \y*4) {};
    \fi
}

% Draw inner dots in the sub-grid
%\foreach \i in {0.5, 1.5, 2.5, 3.5} {
%    \foreach \j in {0.5, 1.5, 2.5, 3.5} {
%        \node[fill=black, circle, scale=0.75] at (1*4+\i, 1*4+\j) {};
%    }
%}

% Add green dots (central points in subgrids)
\foreach \x in {0, 1, 2} {
    \foreach \y in {0, 1, 2} {
        \node[fill=darkgreen, circle, scale=1.1] at (\x*4+2, \y*4+2) {};
    }
}

% Add blue dots (user devices) randomly within each sub-grid
\node[blue, scale=1.75] at (1.7724840612014607, 2.4347701852001378) {\textbf{\texttimes}}; 
\node[blue, scale=1.75] at (2.9829433548228073, 0.924027015315557) {\textbf{\texttimes}}; 
\node[blue, scale=1.75] at (2.644441885892013, 3.0308403296411806) {\textbf{\texttimes}}; 
\node[blue, scale=1.75] at (1.9629436060413763, 0.5942985917601562) {\textbf{\texttimes}}; 
\node[blue, scale=1.75] at (2.8619980222605084, 0.6525879903370426) {\textbf{\texttimes}}; 
\node[blue, scale=1.75] at (0.31143105257260073, 1.8079690484163247) {\textbf{\texttimes}}; 
\node[blue, scale=1.75] at (3.577345917662132, 1.9169360501252743) {\textbf{\texttimes}}; 
\node[blue, scale=1.75] at (2.1461460323921773, 2.0064536963839728) {\textbf{\texttimes}}; 
\node[blue, scale=1.75] at (0.36810655084303123, 3.3040122489665222) {\textbf{\texttimes}}; 
\node[blue, scale=1.75] at (1.0552018808722754, 0.5038747667917374) {\textbf{\texttimes}}; 
\node[blue, scale=1.75] at (2.254918486168064, 0.5519643565652466) {\textbf{\texttimes}}; 
\node[blue, scale=1.75] at (2.93234174900596, 2.8662892566134026) {\textbf{\texttimes}}; 
\node[blue, scale=1.75] at (1.874689852233681, 0.24459590796730213) {\textbf{\texttimes}}; 
\node[blue, scale=1.75] at (2.4120443334421524, 1.0286565327634443) {\textbf{\texttimes}}; 
\node[blue, scale=1.75] at (1.5586603585807497, 3.0626657311184946) {\textbf{\texttimes}}; 
\node[blue, scale=1.75] at (3.5528267124564605, 1.595025092092008) {\textbf{\texttimes}}; 
\node[blue, scale=1.75] at (1.087352422179939, 1.2764896796905147) {\textbf{\texttimes}}; 
\node[blue, scale=1.75] at (2.6431235338178536, 3.0904465673714427) {\textbf{\texttimes}}; 
\node[blue, scale=1.75] at (3.4260966214236874, 0.5596383354660186) {\textbf{\texttimes}}; 
\node[blue, scale=1.75] at (1.9701109844456832, 1.5198477569752717) {\textbf{\texttimes}}; 
\node[blue, scale=1.75] at (1.790777380719999, 6.042992990928731) {\textbf{\texttimes}}; 
\node[blue, scale=1.75] at (0.019539414477018013, 4.436723607987652) {\textbf{\texttimes}}; 
\node[blue, scale=1.75] at (3.2970759064120507, 5.9745885483007) {\textbf{\texttimes}}; 
\node[blue, scale=1.75] at (2.5588162157130214, 6.019481964007415) {\textbf{\texttimes}}; 
\node[blue, scale=1.75] at (1.7148617687894379, 6.85202595329867) {\textbf{\texttimes}}; 
\node[blue, scale=1.75] at (2.609992808601268, 5.9761574695011195) {\textbf{\texttimes}}; 
\node[blue, scale=1.75] at (3.092725661222082, 6.721421733477111) {\textbf{\texttimes}}; 
\node[blue, scale=1.75] at (1.0469678791891495, 4.684141812839158) {\textbf{\texttimes}}; 
\node[blue, scale=1.75] at (0.46265780930426903, 6.082619938210552) {\textbf{\texttimes}}; 
\node[blue, scale=1.75] at (3.392033194494975, 7.256199791715529) {\textbf{\texttimes}}; 
\node[blue, scale=1.75] at (0.6482612922494289, 4.27326191187554) {\textbf{\texttimes}}; 
\node[blue, scale=1.75] at (0.6213266497839061, 4.8718981837599085) {\textbf{\texttimes}}; 
\node[blue, scale=1.75] at (3.1765366015958243, 6.419133237708122) {\textbf{\texttimes}}; 
\node[blue, scale=1.75] at (2.888434375267841, 6.178424155084704) {\textbf{\texttimes}}; 
\node[blue, scale=1.75] at (1.3153302643553237, 7.468820307467853) {\textbf{\texttimes}}; 
\node[blue, scale=1.75] at (0.4103459126814828, 4.5479922561481825) {\textbf{\texttimes}}; 
\node[blue, scale=1.75] at (0.4494860122800106, 6.7153192159501) {\textbf{\texttimes}}; 
\node[blue, scale=1.75] at (3.390173247286337, 4.339599296777276) {\textbf{\texttimes}}; 
\node[blue, scale=1.75] at (3.4542096544774767, 5.2209289536129955) {\textbf{\texttimes}}; 
\node[blue, scale=1.75] at (2.111287802391666, 4.997412903814515) {\textbf{\texttimes}}; 
\node[blue, scale=1.75] at (3.2433970053665107, 8.14496142343432) {\textbf{\texttimes}}; 
\node[blue, scale=1.75] at (2.3057973929378277, 10.533644079181032) {\textbf{\texttimes}}; 
\node[blue, scale=1.75] at (0.881380433338876, 10.010904833987608) {\textbf{\texttimes}}; 
\node[blue, scale=1.75] at (2.3456361945784896, 10.209929121057733) {\textbf{\texttimes}}; 
\node[blue, scale=1.75] at (1.0695988565510635, 9.335404865086097) {\textbf{\texttimes}}; 
\node[blue, scale=1.75] at (3.4204118719843573, 11.10243263436721) {\textbf{\texttimes}}; 
\node[blue, scale=1.75] at (3.8460282737600866, 8.79369845551648) {\textbf{\texttimes}}; 
\node[blue, scale=1.75] at (1.8270615412941993, 11.51205077346907) {\textbf{\texttimes}}; 
\node[blue, scale=1.75] at (1.046113544530355, 8.83178915506555) {\textbf{\texttimes}}; 
\node[blue, scale=1.75] at (2.6024466427304183, 8.730363159423998) {\textbf{\texttimes}}; 
\node[blue, scale=1.75] at (0.154792411222485, 10.956193701650284) {\textbf{\texttimes}}; 
\node[blue, scale=1.75] at (3.9185180271559057, 11.659802322160086) {\textbf{\texttimes}}; 
\node[blue, scale=1.75] at (0.012222560379396619, 9.252605821450041) {\textbf{\texttimes}}; 
\node[blue, scale=1.75] at (2.5342963858591943, 10.613211847988214) {\textbf{\texttimes}}; 
\node[blue, scale=1.75] at (0.8197149643833193, 8.91008797994229) {\textbf{\texttimes}}; 
\node[blue, scale=1.75] at (1.038633989527085, 10.087646396764168) {\textbf{\texttimes}}; 
\node[blue, scale=1.75] at (1.4101526562791662, 9.337151790266567) {\textbf{\texttimes}}; 
\node[blue, scale=1.75] at (0.6863181457723355, 9.595497016367831) {\textbf{\texttimes}}; 
\node[blue, scale=1.75] at (2.7504082379701282, 11.468342270588078) {\textbf{\texttimes}}; 
\node[blue, scale=1.75] at (3.260539554345051, 11.79924071107122) {\textbf{\texttimes}}; 
\node[blue, scale=1.75] at (7.345200268624305, 0.4013130165968697) {\textbf{\texttimes}}; 
\node[blue, scale=1.75] at (5.958553013906176, 2.992957512976153) {\textbf{\texttimes}}; 
\node[blue, scale=1.75] at (4.704498848396732, 0.2368220650115167) {\textbf{\texttimes}}; 
\node[blue, scale=1.75] at (5.946399170599417, 0.5321509359665644) {\textbf{\texttimes}}; 
\node[blue, scale=1.75] at (6.862272681411838, 0.6289305309862172) {\textbf{\texttimes}}; 
\node[blue, scale=1.75] at (5.774306139798339, 0.09190052014314443) {\textbf{\texttimes}}; 
\node[blue, scale=1.75] at (5.9884935915997834, 3.9413104600405027) {\textbf{\texttimes}}; 
\node[blue, scale=1.75] at (7.841714762393822, 1.5457129309795894) {\textbf{\texttimes}}; 
\node[blue, scale=1.75] at (4.340792788069862, 1.5182595595427593) {\textbf{\texttimes}}; 
\node[blue, scale=1.75] at (5.001882025648494, 0.6256356876899933) {\textbf{\texttimes}}; 
\node[blue, scale=1.75] at (4.354778713243845, 0.286946822699623) {\textbf{\texttimes}}; 
\node[blue, scale=1.75] at (4.188904750808743, 1.3709006111642439) {\textbf{\texttimes}}; 
\node[blue, scale=1.75] at (7.122392145300594, 3.296130437442145) {\textbf{\texttimes}}; 
\node[blue, scale=1.75] at (7.108172094336212, 0.8590298924853546) {\textbf{\texttimes}}; 
\node[blue, scale=1.75] at (4.606865388802085, 1.1856399774110895) {\textbf{\texttimes}}; 
\node[blue, scale=1.75] at (5.790550124030407, 0.8664913239487584) {\textbf{\texttimes}}; 
\node[blue, scale=1.75] at (4.8542355953619305, 2.030967245957464) {\textbf{\texttimes}}; 
\node[blue, scale=1.75] at (6.171608685758529, 0.0695863578194289) {\textbf{\texttimes}}; 
\node[blue, scale=1.75] at (4.988056570495592, 3.2067142520433047) {\textbf{\texttimes}}; 
\node[blue, scale=1.75] at (6.735072937830388, 1.2844697710006132) {\textbf{\texttimes}}; 
\node[blue, scale=1.75] at (4.8628544029892655, 7.827393724914357) {\textbf{\texttimes}}; 
\node[blue, scale=1.75] at (6.042548439144518, 5.942939376012532) {\textbf{\texttimes}}; 
\node[blue, scale=1.75] at (5.017705155844483, 5.5392722088527915) {\textbf{\texttimes}}; 
\node[blue, scale=1.75] at (7.578664886691596, 7.853158457264829) {\textbf{\texttimes}}; 
\node[blue, scale=1.75] at (5.485822074862378, 4.467670438399355) {\textbf{\texttimes}}; 
\node[blue, scale=1.75] at (5.628518999439773, 4.965260015895896) {\textbf{\texttimes}}; 
\node[blue, scale=1.75] at (7.455520634491775, 7.319083184154491) {\textbf{\texttimes}}; 
\node[blue, scale=1.75] at (4.326852449839981, 7.172760576776229) {\textbf{\texttimes}}; 
\node[blue, scale=1.75] at (4.282540958180595, 6.01571929074809) {\textbf{\texttimes}}; 
\node[blue, scale=1.75] at (7.54398002535813, 7.20741936447633) {\textbf{\texttimes}}; 
\node[blue, scale=1.75] at (6.984367978094824, 7.47863726801509) {\textbf{\texttimes}}; 
\node[blue, scale=1.75] at (4.642901747216206, 6.531050941364013) {\textbf{\texttimes}}; 
\node[blue, scale=1.75] at (4.579174832597747, 5.76116817153543) {\textbf{\texttimes}}; 
\node[blue, scale=1.75] at (5.392506659107214, 6.80608231304761) {\textbf{\texttimes}}; 
\node[blue, scale=1.75] at (6.268294460795541, 5.855781293991743) {\textbf{\texttimes}}; 
\node[blue, scale=1.75] at (5.6331275044327365, 7.046929611737356) {\textbf{\texttimes}}; 
\node[blue, scale=1.75] at (6.133519217237376, 4.972362607393856) {\textbf{\texttimes}}; 
\node[blue, scale=1.75] at (5.797904179196082, 5.393740355685619) {\textbf{\texttimes}}; 
\node[blue, scale=1.75] at (7.592807059876913, 4.349337488581371) {\textbf{\texttimes}}; 
\node[blue, scale=1.75] at (4.725419131331908, 6.082566491255482) {\textbf{\texttimes}}; 
\node[blue, scale=1.75] at (5.615617471555746, 10.141070652987944) {\textbf{\texttimes}}; 
\node[blue, scale=1.75] at (5.258030431449373, 8.988239213051369) {\textbf{\texttimes}}; 
\node[blue, scale=1.75] at (4.960965647232305, 10.749643378323967) {\textbf{\texttimes}}; 
\node[blue, scale=1.75] at (7.686688074166355, 11.87200843853212) {\textbf{\texttimes}}; 
\node[blue, scale=1.75] at (5.123485320547507, 10.552894299991685) {\textbf{\texttimes}}; 
\node[blue, scale=1.75] at (4.482842136046603, 9.654238618342495) {\textbf{\texttimes}}; 
\node[blue, scale=1.75] at (5.864158566651199, 10.44146745429506) {\textbf{\texttimes}}; 
\node[blue, scale=1.75] at (4.7130939614134295, 9.946537084694047) {\textbf{\texttimes}}; 
\node[blue, scale=1.75] at (5.206754804533439, 10.96011746831628) {\textbf{\texttimes}}; 
\node[blue, scale=1.75] at (4.233953383579552, 8.98409163954447) {\textbf{\texttimes}}; 
\node[blue, scale=1.75] at (4.942911139533211, 10.509671318022455) {\textbf{\texttimes}}; 
\node[blue, scale=1.75] at (6.942814458314507, 8.011333427376998) {\textbf{\texttimes}}; 
\node[blue, scale=1.75] at (5.306273094180392, 11.73675700155544) {\textbf{\texttimes}}; 
\node[blue, scale=1.75] at (4.338679486324674, 10.887710080119565) {\textbf{\texttimes}}; 
\node[blue, scale=1.75] at (7.550880320974962, 10.815876674603906) {\textbf{\texttimes}}; 
\node[blue, scale=1.75] at (6.338104025928829, 9.137658555204107) {\textbf{\texttimes}}; 
\node[blue, scale=1.75] at (6.11457217514816, 8.350667889190756) {\textbf{\texttimes}}; 
\node[blue, scale=1.75] at (4.956358950628059, 9.76495640273357) {\textbf{\texttimes}}; 
\node[blue, scale=1.75] at (7.189928521756936, 11.330163067317637) {\textbf{\texttimes}}; 
\node[blue, scale=1.75] at (5.088630721545101, 9.161980949432774) {\textbf{\texttimes}}; 
\node[blue, scale=1.75] at (8.94775913732331, 3.125531461937196) {\textbf{\texttimes}}; 
\node[blue, scale=1.75] at (8.369115462483704, 2.049378840193383) {\textbf{\texttimes}}; 
\node[blue, scale=1.75] at (8.527576643170345, 0.8275638966532708) {\textbf{\texttimes}}; 
\node[blue, scale=1.75] at (11.104166897717523, 1.1176119317356856) {\textbf{\texttimes}}; 
\node[blue, scale=1.75] at (10.68677458500319, 1.0210796180112416) {\textbf{\texttimes}}; 
\node[blue, scale=1.75] at (8.026317145267758, 2.208524648481097) {\textbf{\texttimes}}; 
\node[blue, scale=1.75] at (9.034008496601801, 0.8558152835453701) {\textbf{\texttimes}}; 
\node[blue, scale=1.75] at (11.632041162412754, 1.034761322613552) {\textbf{\texttimes}}; 
\node[blue, scale=1.75] at (8.479244152644108, 1.1667964924456622) {\textbf{\texttimes}}; 
\node[blue, scale=1.75] at (11.781847226151628, 2.4076785204038935) {\textbf{\texttimes}}; 
\node[blue, scale=1.75] at (8.387606220432758, 2.198185563016891) {\textbf{\texttimes}}; 
\node[blue, scale=1.75] at (9.063289571457585, 1.1849233763370783) {\textbf{\texttimes}}; 
\node[blue, scale=1.75] at (11.911021614718742, 2.997282182398302) {\textbf{\texttimes}}; 
\node[blue, scale=1.75] at (8.359000228270146, 2.4421124272142882) {\textbf{\texttimes}}; 
\node[blue, scale=1.75] at (10.891187610260413, 0.5194039544653162) {\textbf{\texttimes}}; 
\node[blue, scale=1.75] at (10.859240161868161, 2.8500618869846384) {\textbf{\texttimes}}; 
\node[blue, scale=1.75] at (10.484931349824407, 3.974002624220472) {\textbf{\texttimes}}; 
\node[blue, scale=1.75] at (10.810930783977197, 2.7241333755687847) {\textbf{\texttimes}}; 
\node[blue, scale=1.75] at (11.625220238495908, 3.146334282127714) {\textbf{\texttimes}}; 
\node[blue, scale=1.75] at (11.74503726456395, 0.9758461650090933) {\textbf{\texttimes}}; 
\node[blue, scale=1.75] at (11.94187568333326, 5.844619138432524) {\textbf{\texttimes}}; 
\node[blue, scale=1.75] at (9.10791922034176, 7.743241701337044) {\textbf{\texttimes}}; 
\node[blue, scale=1.75] at (11.714769097083783, 6.497314247282317) {\textbf{\texttimes}}; 
\node[blue, scale=1.75] at (9.790402855117158, 6.7115178557654165) {\textbf{\texttimes}}; 
\node[blue, scale=1.75] at (10.738277927211175, 5.0421947937299425) {\textbf{\texttimes}}; 
\node[blue, scale=1.75] at (11.486037926641316, 6.992011697729099) {\textbf{\texttimes}}; 
\node[blue, scale=1.75] at (11.384908060667932, 4.268492976040406) {\textbf{\texttimes}}; 
\node[blue, scale=1.75] at (8.81169850007265, 7.485618123004598) {\textbf{\texttimes}}; 
\node[blue, scale=1.75] at (8.294656661743346, 7.094452799462043) {\textbf{\texttimes}}; 
\node[blue, scale=1.75] at (9.48296697050033, 7.243448659514957) {\textbf{\texttimes}}; 
\node[blue, scale=1.75] at (8.101865382443531, 5.305435766465363) {\textbf{\texttimes}}; 
\node[blue, scale=1.75] at (9.716099047427353, 4.88547978240277) {\textbf{\texttimes}}; 
\node[blue, scale=1.75] at (10.315268109158037, 5.950733821093639) {\textbf{\texttimes}}; 
\node[blue, scale=1.75] at (10.92194408968118, 7.4774461876933875) {\textbf{\texttimes}}; 
\node[blue, scale=1.75] at (10.992442384979626, 4.680703781757033) {\textbf{\texttimes}}; 
\node[blue, scale=1.75] at (9.897328034788831, 5.482504518228996) {\textbf{\texttimes}}; 
\node[blue, scale=1.75] at (10.769650710689616, 6.962602375082078) {\textbf{\texttimes}}; 
\node[blue, scale=1.75] at (8.175910596731242, 6.826312724407893) {\textbf{\texttimes}}; 
\node[blue, scale=1.75] at (11.679256836819734, 7.809515416408189) {\textbf{\texttimes}}; 
\node[blue, scale=1.75] at (10.066438311659144, 6.937451886986404) {\textbf{\texttimes}}; 
\node[blue, scale=1.75] at (8.535645568588746, 8.75582932058716) {\textbf{\texttimes}}; 
\node[blue, scale=1.75] at (8.691836016411289, 8.608821106466149) {\textbf{\texttimes}}; 
\node[blue, scale=1.75] at (9.562649024488712, 11.346856776981122) {\textbf{\texttimes}}; 
\node[blue, scale=1.75] at (11.233363894482045, 10.192936554661896) {\textbf{\texttimes}}; 
\node[blue, scale=1.75] at (10.096724794667779, 11.367495342868946) {\textbf{\texttimes}}; 
\node[blue, scale=1.75] at (9.471054868376301, 11.623547899077927) {\textbf{\texttimes}}; 
\node[blue, scale=1.75] at (10.862486129932643, 10.743651860347578) {\textbf{\texttimes}}; 
\node[blue, scale=1.75] at (10.066101990903324, 10.22782644407458) {\textbf{\texttimes}}; 
\node[blue, scale=1.75] at (9.57112815448737, 10.1820965762512) {\textbf{\texttimes}}; 
\node[blue, scale=1.75] at (10.802054772264826, 11.483516632774755) {\textbf{\texttimes}}; 
\node[blue, scale=1.75] at (11.72757719032065, 11.508927427082229) {\textbf{\texttimes}}; 
\node[blue, scale=1.75] at (9.048534585069547, 11.305724442421587) {\textbf{\texttimes}}; 
\node[blue, scale=1.75] at (9.16920422992326, 8.021922417348751) {\textbf{\texttimes}}; 
\node[blue, scale=1.75] at (11.323621203945558, 10.795640774869945) {\textbf{\texttimes}}; 
\node[blue, scale=1.75] at (11.545879937002177, 8.901321370979677) {\textbf{\texttimes}}; 
\node[blue, scale=1.75] at (10.653920189244044, 11.194100833693586) {\textbf{\texttimes}}; 
\node[blue, scale=1.75] at (10.29968827721157, 9.903945764334413) {\textbf{\texttimes}}; 
\node[blue, scale=1.75] at (11.816509905005061, 8.818386723587647) {\textbf{\texttimes}}; 
\node[blue, scale=1.75] at (10.301628291431074, 8.523264686801365) {\textbf{\texttimes}}; 
\node[blue, scale=1.75] at (11.104229940002776, 11.073213157288905) {\textbf{\texttimes}};

% Draw legend on the right
%\draw[thin, rounded corners] (13, 4.2) rectangle (19, 7.6);
\node[fill=yelloworange, rectangle, scale=2] at (13.75, 7.24) {};
\tikzset{legend text/.style={font=\huge}}
\node[legend text, anchor=west, scale=1.3] at (14.4, 7.2) {access point};
\node[fill=darkgreen, circle, scale=1.1] at (13.75, 6.0) {};
\node[legend text, anchor=west, scale=1.3] at (14.4, 6.0) {zone centroid};
\node[blue, scale=1.75] at (13.75, 4.77) {\textbf{\texttimes}};
\node[legend text, anchor=west, scale=1.3] at (14.4, 4.8) {user device};

\end{tikzpicture}

%% file: figs/block_diagram.tex
\begin{tikzpicture}[node distance=0.5cm and 0.7cm, font=\scriptsize]

    % Quant block with incoming arrow for sensed message input
    %\node[draw, rectangle, minimum width=0.9cm, minimum height=0.6cm] (quant) {Quantizer $\text{quant}_{\setQ_u}$};
    %\node[left=1.1cm of quant, circle, fill, inner sep=0.8pt] (input) {};
    %\node[left=0.6cm of quant, inner sep=0.8pt] (input) {};
    %\draw[->] (input) -- node[above, xshift=-1.5mm] {\scriptsize $S_{u,k}$} (quant.west);
    
    % Enc block (smaller)
    \node[draw, rectangle, 
    %right= 1cm of quant, 
    minimum width=0.9cm, minimum height=0.6cm] (enc) {Encoder $\enc_{u}$};
    \node[left=0.3cm of enc, inner sep=0.8pt] (input) {$W_{u,k}$};
    \draw[->] (input) -- (enc.west);
    %\draw[->] (input) -- node[above, xshift=-1.5mm] {\scriptsize $W_{u,k}$} (enc.west);

    % Multiplier block with sqrt(NP)
    \node[draw, circle, right=1.35cm of enc, minimum size=0.1cm] (mult) {$\times$};
    \node[above=0.2cm of mult] (np) {};
    \node[above=0.15cm of mult] (np_text) {\scriptsize $\sqrt{\uN\uP}$};
    
    % Shortened vertical arrow from NP to multiplier
    %\draw[->] (np) -- ++(0,-0.2) -- (mult.north);

    % Channel vector h_{u,m,k}
    %\node[draw, rectangle, right=0.3cm of mult, minimum width=0.9cm, minimum height=0.4cm] (channel) {$\vech_{u,k}$};
    \node[draw, circle, right=0.3cm of mult, minimum size=0.1cm] (channel_mult) {$\times$};
    \node[above=0.2cm of channel_mult] (channel_label) {};
    \node[above=0.15cm of channel_mult] (channel_label_text) {\scriptsize $\tp{\vech_{u,k}}$};
    
    % Summation block for W, aligned horizontally
    \node[draw, circle, right=0.3cm of channel_mult, minimum size=0.4cm] (adder) {$\sum$};
    \node[draw, circle, right=0.3cm of adder, minimum size=0.4cm] (adder2) {$\sum$};
    \node[above right=0.1cm of adder2.north east] (w_source_below) {};
    \node[above=0.5cm of w_source_below.center] (w_source) {\scriptsize $\mathbf{W}$};

    \node[below right=0.1cm of adder2.south east] (y_source_above) {};
    \node[below=2.2cm of y_source_above.center] (y_source) {%\scriptsize $\mathbf{Y}$
    };

    % Shortened vertical arrow from W to sum
    \draw (w_source) -- ++(0,-0.25) -- (w_source_below.center);
    \draw[->] (w_source_below.center) -- (adder2.north east);
    \draw (y_source.north) -- (y_source_above.center);
    \draw (y_source_above.center) -- (adder2.south east);
    
    % Connections for Zone u
    %\draw[->] (quant.east) -- node[above] {\scriptsize $W_{u,k}$} (enc.west);
    \draw[->] (enc) -- node[above] {\scriptsize $\vecc_{u,W_{u,k}}$} (mult);
    \draw[->] (np) -- (mult); 
    \draw[->] (mult) -- (channel_mult);

    \draw[->] (channel_label) -- (channel_mult);
    \draw[->] (channel_mult) -- (adder);
    \draw[->] (adder) -- (adder2);

    % Output to Y
    \node[draw, rectangle, left = 2.7cm of y_source.north, minimum width=0.9cm, minimum height=0.6cm] (dec) {Decoder $\dec$}; % 1.3cm
    \node[left= 0.6cm of dec, minimum width=0.4cm, inner sep=1.8pt] (multiplicity) { $\displaystyle 
    \widehat{\vect}    %\sum_{u=1}^{\uU}\sum_{m=1}^{\uM_u} \widehat{t}_{u,m} \delta_{u,m} %(\vecq_{u,m})
    $}; 

    \draw[->] (y_source.north) -- node[above] {\scriptsize $\mathbf{Y}$} (dec);
    \draw[->] (dec) -- (multiplicity);

    % Zone u Label inside shaded area, near the top-left corner
    \node[above right = 0.4cm and -0.85cm of input, 
    red] (user_label) { user $k$ in zone $u$};
    \node[above right=1.27cm and -0.9cm of input, 
    black] {zone $u$}; % {Zone $u$};
    %\node[above right = 1.5cm and -0.33cm of input, black] {Zone $u$};

    %\node[above left = 0cm and 0.35cm of np] (upvdots) {$\vdots$};
    %\node[below left = 0.8cm and 0.35cm of np] (downvdots) {$\vdots$};

    \node[above = 1.5cm of adder.center] (upuserstart) {}; %channel.east
    \node[below = 0.95cm of adder.center] (downuserstart) {};

    %\node[draw, rectangle, left=0.56cm of upuserstart, minimum width=0.9cm, minimum height=0.4cm] (channel1) {$\vech_{u,m,k}$};

    \node[left = 4.5cm of upuserstart, anchor=west] (leftupuserstart) {From user $1$ in zone $u$};
    \node[below right = -0.045cm and 0.9cm of leftupuserstart.west] (upvdots) {\scalebox{0.8}{$\vdots$}};
    \node[left = 4.5cm of downuserstart, anchor=west] (leftdownuserstart) {From user $\mathrm{K}_{u}$ in zone $u$};
    \node[above right = 0.09cm and 0.9cm of leftdownuserstart.west] (downvdots) {\scalebox{0.8}{$\vdots$}};

    \draw (upuserstart.center) -- (leftupuserstart.east);
    \draw (downuserstart.center) -- (leftdownuserstart.east);
    \draw[->] (upuserstart.center) -- (adder);
    \draw[->] (downuserstart.center) -- (adder);

    \node[above = 2.5cm of adder2.center] (upzonestart) {}; % adder.east
    \node[below = 1.9cm of adder2.center] (downzonestart) {}; %adder.east % old: 1.8cm

    \draw[->] (upzonestart.center) -- (adder2);
    \draw[->] (downzonestart.center) -- (adder2);

    \node[left = 3cm of upzonestart, anchor=west] (leftupzonestart) {From zone $1$};
    \node[below right = -0.01cm and 0.9cm of leftupzonestart.west] (upvdots2) {\scalebox{0.8}{$\vdots$}};
    \node[left = 3cm of downzonestart, anchor=west] (leftdownzonestart) {From zone $\uU$};
    \node[above right = 0.1cm and 0.9cm of leftdownzonestart.west] (downvdots2) {\scalebox{0.8}{$\vdots$}};

    \draw (upzonestart.center) -- (leftupzonestart.east);
    \draw (downzonestart.center) -- (leftdownzonestart.east);

    % Background shading for Zone u
    \begin{scope}[on background layer]
        \node[draw=none, fill=gray!10, rounded corners, fit={%(quant) 
        (input) (enc) (channel_mult) (upvdots) (downvdots) (adder) (leftupuserstart) (leftdownuserstart)}, inner xsep=0.3cm, inner ysep=0.1cm, yshift=-0cm, xshift=-0.19cm] (zoneu) {};
        \node[draw, dashed, red, rounded corners, fit={%(quant) 
        (input) (enc) (user_label) (channel_mult) (channel_label) }, inner xsep=0.15cm, inner ysep=0.1cm, yshift=-0.05cm, xshift=-0.0cm] (zoneu) {};
    \end{scope}
\end{tikzpicture}